# The Persistence of Memory in Solid-State Ionic Conduction Probed by Nonlinear Optics

Andrey D. Poletayev[1,2,3,*], Matthias C. Hoffmann[4], James A. Dawson[5,6], Samuel W. Teitelbaum[1,7,8], Mariano Trigo[1,7], M. Saiful Islam[3,9], Aaron M. Lindenberg[1,2,7,*]

**Predicting practical rates of ion transport from atomistic descriptors enables the rational design of materials, devices, and processes, which is especially critical to developing low-carbon energy technologies such as rechargeable batteries[1–3]. The correlated mechanisms of ionic conduction[4,5], variation of conductivity with timescale[6–8] and confinement[9,10], and ambiguity in the vibrational origin of translation, the attempt frequency[11,12], call for a direct atomic probe of the most fundamental steps of ionic diffusion: ion hops. However, such hops are rare-event large-amplitude translations, and are challenging to excite and detect. Here we use single-cycle terahertz pumps to impulsively trigger ionic hopping in battery solid electrolytes. This is visualized by an induced transient birefringence enabling direct probing of anisotropy in ionic hopping on the picosecond timescale. The relaxation of the transient signal measures the decay of orientational memory, and the production of entropy in diffusion. We extend experimental results using *in silico* transient birefringence to identify attempt frequencies for ion hopping. Using nonlinear optical methods, we probe ion transport at its fastest limit, distinguish correlated conduction mechanisms from a true random walk at the atomic scale, and demonstrate the connection between activated transport and the thermodynamics of information.**

## Models of Ion Transport

Fast-ion conduction in the solid state commands particular attention due to its importance in energy technologies, such as solid state batteries[1,2]. Linking the mechanistic features of ion conduction at the atomic level to the collective descriptors of macroscopic transport within a multi-scale model, or as measured in a device, yields opportunities to design novel processes and applications[3,9,13]. However, the links are often not well characterized in part because nanoscale ionic conductivity is dispersive and non-ergodic[6,7].

Macroscopic solid-state ionic conduction is considered a random-walk process composed of rapid translations between lattice sites. These fundamental steps of diffusion and conduction are called ion hops (Figure 1a). Hops originate randomly with an attempt frequency $v_0$, and succeed with a probability determined by the Gibbs free energy of a transition state[6,11,12,14]. Random-walk models[15,16] treat every step (hop) as uncorrelated from the preceding ones:

[1] Stanford Institute for Materials and Energy Sciences, SLAC National Laboratory, Menlo Park, CA, USA
[2] Department of Materials Science and Engineering, Stanford University, Stanford, CA, USA
[3] Department of Materials, University of Oxford, Oxford, UK
[4] Linac Coherent Light Source, SLAC National Accelerator Laboratory, Menlo Park, CA, USA
[5] Chemistry – School of Natural and Environmental Sciences, Newcastle University, Newcastle upon Tyne, UK
[6] Centre for Energy, Newcastle University, Newcastle upon Tyne, UK
[7] Stanford PULSE Institute, SLAC National Accelerator Laboratory, Menlo Park, CA 94025, USA
[8] Present Address: Department of Physics, Arizona State University, Tempe, AZ 85287, USA
[9] Department of Chemistry, University of Bath, Bath, UK
[*] Correspondence to andrey.poletayev@gmail.com, aaronl@stanford.edu.

all memory is lost with every hop. From the standpoint of information theory[17], the full entropy of conduction is evolved at every hop.

By contrast, models of correlated transport allow for the macroscopic process of conduction to consist of multiple interconnected steps. Examples include memory kernels in generalized master equations[18,19], Burnett-order nonlinear hydrodynamics[8,10], or kinetic competition[6,20]. In such models, the state of the material at time zero (Figure 1b, green) influences transport dynamics over some non-negligible timescale (e.g., $t_1$ in Figure 1b). The information entropy of transport is expressed as a mutual entropy between configurations a time lag $t_1$ apart. This quantity increases (equivalently, mutual information decreases) until the full entropy of transport is produced. Only at longer timescales (e.g., at $t_2$ in Figure 1b) can transport parameters such as the diffusion coefficient and ionic conductivity be constant-valued, as in a random-walk process. Indeed, picosecond to nanosecond studies of ion conductors[11,21,22] yield reduced activation energies, suggesting that the processes being probed at those frequencies may be incomplete with respect to overall conduction due to the persistence of such memory.

Understanding correlation effects in transport remains necessary to predict practical performance of ionic conductors from the atoms up[3,13], and to exploit nonlinear nanoscale transport phenomena[9] in devices. Here, we study the correspondence between the thermodynamics of ion transport and those of information[23,24]. Such a mechanistic investigation requires the ability to (1) impulsively trigger ion hops[3] on the short timescale of $1/v_0$, typically ≤ 1 ps, and (2) track their outcomes over potentially much longer times[7,8]. Such a combination is inaccessible to current techniques[11], and even the vibrational nature of $v_0$ remains under debate[2,25,26]. Furthermore, since ion hops, unlike those of electronic carriers, are stochastic and lack spectral resonance signatures, single-pulse reciprocal-space or Fourier-transform probes are insufficient for probing them. Instead, correlations in nanoscale transport[7-10] could be probed using time-domain nonlinear optical pump-probe methods such as transient birefringence[27-29] via their sensitivity to polarization. However, pump-probe studies of ionic dynamics have so far focused on coherent displacements of bound vibrational modes or order parameters[30,31], where many ions undergo coherent sub-angstrom motions, rather than rare large-amplitude displacements (typically 2-3 Å) such as ionic hops.

Probing Correlated Hopping

We probe orientational correlations between ionic hops using a nonlinear optical method sensitive to ionic *velocities* (Figure 1c). We trigger ionic hops with an impulsive single-optical-cycle pump (Figure 1d) and measure the anisotropy in the directions of subsequent hops (Figure 1ef) as transient birefringence. The ionic response is pumped impulsively using single-cycle terahertz pulses with center frequency ≈0.7 THz[32,33], and transient birefringence is probed in transmission mode (Extended Data Figure 1). The general solution for electric fields $E$ in the sample at position $z$ along the pulse paths at pump-probe delay $t_1$ is[34,35]

$$E_{probe}(z,t) - E(z,t) \propto \frac{\partial}{\partial t}\Big(P(z, t+t_1)E_{probe}(z,t)\Big)$$

Here, the polarization $P$ arises from coherent displacements of vibrational modes or bound dipoles $Q_i$, plus an additional component from the history of hopping rates $H$:

$$P(t + t_1) = \sum_i Q_i(t + t_1) + \int_{-\infty}^{t+t_1} H(\tau)d\tau$$

The integral is zero until the pump drives ionic hops in a preferential direction (Figure 1d). Because the emitted signal at $t_1$ is proportional to the time derivative of the polarization, it arises from ionic velocities, to which hopping *rates $H(t_1)$* contribute. In the terahertz-pumped Kerr effect (TKE) geometry, the *anisotropy* of velocities and hopping rates is measured specifically (Figure 1c). This corresponds to a time-dependent *preference* for hopping along the direction defined by the terahertz pump versus in an orthogonal direction. The relaxation of hopping anisotropy corresponds to the loss of memory of the pump-driven impulse by the conducting ions during the temperature-activated solid-state diffusion process. If conduction occurs via a random walk, then hopping immediately returns to isotropy, and no relaxation should be observable. Additionally, the modes $Q_i$ that couple appreciably to $H$ contribute to attempt frequencies $\nu_0$.

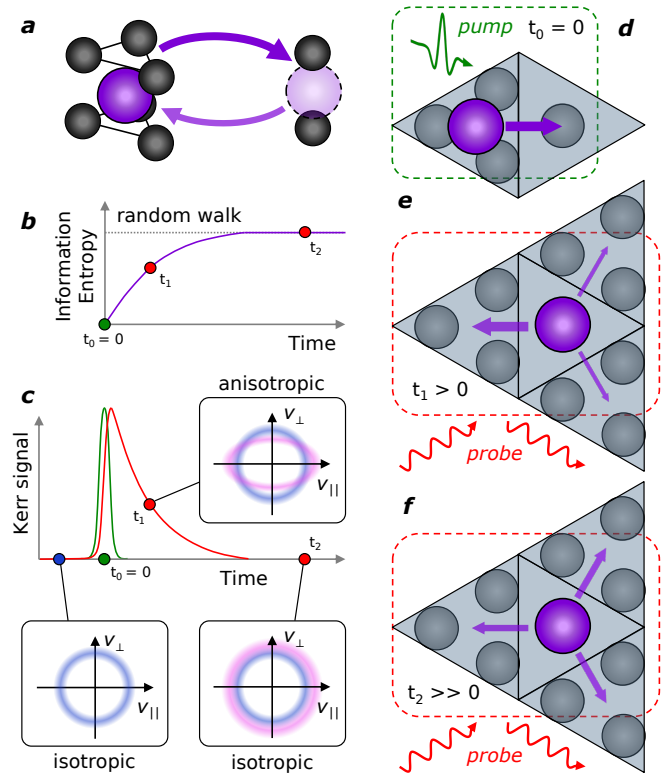

**Figure 1 | Probing Ionic Transport in *β*-aluminas**. **a**. Mobile metal ions (purple) hop (arrows) between energetically distinct lattice sites with two- and six-fold coordination by oxygen ions (black). The same lattice sites are depicted schematically in **d-f**. **b**. At time $t_0 = 0$ (green) there is perfect knowledge of the state of the system. At time $t_1 > 0$ some correlations persist. By time $t_2 \gg 0$ all correlations and memory are lost, and the full entropy of transport is produced as for a random walk (dotted line). **c**. Transient birefringence experiment in *β*-aluminas: At time $t_0 = 0$ (green) the terahertz pump triggers ion hops along its electric field vector, as shown in **d**. At time $t_1 > 0$, the subsequent hops retain a correlation with the impulse at $t_0$: back-hopping is predominant, as shown in **e**, which yields a birefringence signal (**c**, red). The anisotropic distribution of hopping directions at $t_1$ is probed as a contribution to the anisotropy in ionic velocities (**c**, pink). The pump energy fully thermalizes by time $t_2$ when the hopping directions reach isotropy (**f**).

## Picosecond Hopping Dynamics in β-aluminas

We first use the fast ionic conductors β-aluminas ($M_{1+2x}Al_{11}O_{17+x}$, where $x \approx 0.1$, and the mobile ion $M^+$ = $Na^+$, $K^+$, $Ag^+$) as model systems. In β-aluminas, ion conduction occurs over two non-equivalent lattice sites (Figure 1a), and correlations persist over timescales corresponding to multiple (≥2) consecutive hops[7]. The vibrational modes of the mobile ions fall between 0.7-3.0 THz, depending on the mobile ion[36,37]. To match the direction of ionic hopping, the pump electric field $E$ is perpendicular to the crystalline $c$ axis and parallel to the two-dimensional conduction planes. Figure 2a shows the time traces of the terahertz-pumped transient birefringence in β-aluminas at 300 K. All samples show both oscillatory and non-oscillatory responses at 300 K and 620 K (Extended Data Figure 1). Such incoherent, non-oscillatory relaxation has been previously observed in liquids[27,29,38,39] and solids[40], but attributed to overdamped rotations or librations, which are absent in β-aluminas. While the magnitude of the birefringence response scales with the square of the pump field (Figure S3), as expected for a Kerr effect signal, we do not observe any spectral changes with increasing pump fluence in either TKE or terahertz transmission experiments (Figures S3, S5), indicating that we are probing a response intrinsic to the material, and not a threshold-dependent response only relevant at high fields[41]. The non-oscillatory relaxation is substantially slower than any vibrational components. Therefore, we also rule out nonlinear phonon coupling[42,43]. Otherwise, the vibrational components of the TKE signals (Figure 2b) are consistent with established far-infrared and Raman modes[37] of mobile ions.

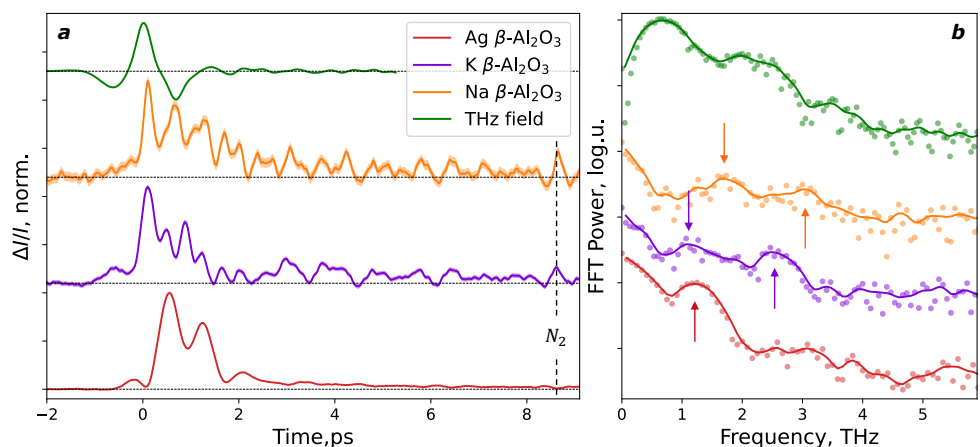

**Figure 2 | Terahertz Kerr effect (TKE) in β-alumina ion conductors.** (**a**) Time-domain traces of transient birefringence in β-alumina ion conductors: $Na^+$ (orange), $K^+$ (purple), $Ag^+$ (dark red), and electro-optic sampling trace of the pump field (green), smoothed with a 30-fs Gaussian window. The shaded regions correspond to ±1 s.e. of the mean signal at each time delay. The labeled feature at 8.4 ps is the rotational coherence of atmospheric nitrogen[44] within a Rayleigh length (> 500 μm) of the samples (≤ 30 μm thick). (**b**) Fourier transform power spectra (points) of the signals in (**a**), and smoothed with a Gaussian filter of 0.1 THz st. dev. (lines), with peaks highlighted by arrows.

We investigate the possible connection between temperature-activated hopping conduction and the picosecond TKE relaxation by varying the temperature of the samples. In all materials, the signals from thinner (<30 μm) vs thicker (100-300 μm) samples match after the short-time oscillations dephase (Extended Data Figure 2). The subsequent non-oscillatory relaxation represents a bulk material response. We analyze a thick sample of K+ β-alumina (Figure 3a) here to highlight this long-time signal. We model the response as the sum of an instantaneous response arising from intrinsic nonlinearity and a mismatch in optical constants between the pump and probe frequencies[45,46] (Supplementary Notes 1-2, and Extended Data Figures 3-5), and single-exponential decay, shown together as dashed lines. The slow non-oscillatory component is absent in the non-resonant optical Kerr response (Extended Data Figure 6), and the residuals of the TKE fit (Extended Data Figure 7a) show the same frequency, ≈2 THz, as the non-resonant signal. Together, this suggests that both pumps (THz and optical) excite coherent ionic vibrations, but the terahertz pump excites an additional response that decays incoherently. For K+ β-alumina, this non-oscillatory relaxation accelerates from ≈10 ps at 300 K to ≈4 ps at 620 K (Figure 3b), with an activation energy of ≈45 meV. In ambient atmosphere, the apparent time constants are slightly slower due to the overlaid signal from atmospheric water vapor[27] (Figure S1), which is slower than the sample response and is not temperature-activated.

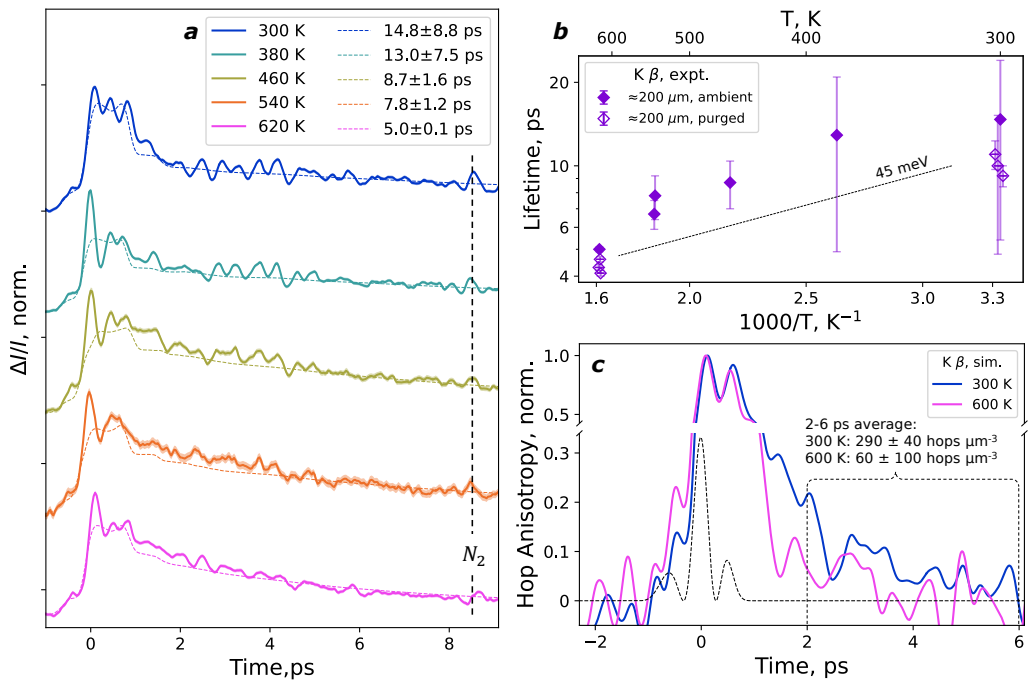

**Figure 3 | Temperature dependence of the long-lived TKE in K+ *β*-alumina.** (**a**) Time-domain traces of transient birefringence in K+ *β*-alumina, measured in ambient atmosphere (solid lines), normalized and offset for clarity. The shaded regions correspond to ±1 s.e. of the mean signal. Colored dashed lines are fits to the sum of single-exponent relaxation and an instantaneous polarization. The labeled feature at 8.4 picoseconds is the rotational coherence of atmospheric nitrogen. (**b**) Time constants of single-exponential fits to the long-lived TKE component in ambient atmosphere (filled symbols) and dry atmosphere (empty symbols). Error bars are ±1 s.e. of least-squares fitting. The dashed line at an activation energy of 45 meV is a guide to the eye. (**c**) Molecular dynamics simulation showing normalized anisotropy of hopping under applied electric field mimicking the experimental terahertz pump at 300 K (blue) and 600 K (pink). Black dashed line: square of the simulated electric field.

Eliminating several other plausible assignments, we propose that the liquid-like picosecond-timescale component of the TKE response arises from the incoherent hopping of mobile ions.

We use large-scale molecular dynamics simulations with a pulsed electrical field[27,47,48] mimicking the experimental THz pump in frequency and magnitude to verify this hypothesis. Following the simulated terahertz pump, the anisotropy of hopping rates is calculated from the times and crystallographic directions of all hops (Figure 3c). The pump selectively accelerates mobile ions (Supplementary Note 4) and drives hopping along its electric field. In K$^+$ $\beta$-alumina at 300 K, the anisotropy of hopping is distinguishable from zero until ≥6 ps after the peak applied field (Figure 3c, blue), but relaxes to zero by ≈3-4 ps at 600 K, in agreement with experiment. The simulated relaxation of the hopping anisotropy is only slightly faster than the experimental TKE relaxation. We conclude that the liquid-like picosecond-timescale TKE response in $\beta$-aluminas is indeed a signature of anisotropy in the hopping of mobile ions caused by the pump, the decay of which is heat-activated with an energy of ≈40-50 meV in K$^+$ $\beta$-alumina (Figure 3b).

Correlations and Memory in Ion Hopping
We now discuss the physical meaning of the TKE signal. The contribution of hopping to the TKE signal for any time delay scales with the hopping rate $H$, weighed by the hop directions relative to the pump's electric field. Hopping rates both parallel (+$v_{//}$) and antiparallel (-$v_{//}$) to the pump contribute positively to the TKE signal, whereas hopping orthogonal to the pump contributes negatively as ±$v_\perp$ (Figure 1c). The *existence* of the slow TKE signal signifies a mismatch between $v_{//}$ and $v_\perp$, and implies that directions of consecutive hops are correlated. This is consistent with previous simulations[6,7]: following pump-driven hops at $t_0$ (Figure 1d), the mobile ions have a preference to hop backwards (correlation factors $f, f_I$ << 1 using solid-state ionics nomenclature[15,16]), which randomizes only slowly. TKE measures the decay of this preference and verifies that the atomistic mechanism of ion transport in $\beta$-aluminas differs from a random walk.

The hopping component of the TKE signal is temperature-activated, but the activation is much smaller than for low-frequency conductivity (≈45 meV in Figure 3b vs >200 meV[49]). However, any hop in isolation is not timescale-dependent. The difference between hopping at picosecond versus macroscopic timescales is not within individual hops, but in the mutual information between successive hops (Supplementary Note 5), conceptually similar to a viscoelastic effect. Small activation energies measured with TKE are consistent with the slow loss of such correlation via production of entropy (Figure 1b), being the origin of increasing measured activation energies toward macroscopic timescales[22]. At picosecond timescales, as probed here, even though hopping is triggered, correlations persist throughout the measurement, and only partial activation is measurable. Conversely, the full activation for macroscopic conduction is measured only at timescales sufficient for the decay of all correlations.

Random-Walk Ion Conduction in K$^+$ $\beta$"-alumina
All $\beta$-aluminas exhibit the picosecond "tail" in the TKE response, and the hopping remains correlated for longer than the TKE response is measurable[7]. This is expected due to the presence of non-equivalent lattice sites (Figure 1). To verify the above interpretation, we next seek a control system where ion transport could proceed via a true atomistic random walk. In such a system, any TKE "tail" could be expected to vanish rapidly. In the closely analogous $\beta$"-aluminas ($M_{1.67}Al_{10.67}Li_{0.33}O_{17}$, where the mobile ion M$^+$ = Na$^+$, K$^+$, Ag$^+$) all lattice sites for ion hopping are equivalent (Figure 4a, inset). The $\beta$"-aluminas are expected to have true random-walk conduction at elevated temperature, but not at 300 K[7]. In the TKE response of K$^+$ $\beta$"-alumina, we observe a long-lived non-oscillatory relaxation at 300 K (Figure 4a), as in the $\beta$-aluminas. At 620 K, this component is absent, and a more field-following response is observed, which suggests the absence of long-time correlations, and a more field-following response overall. In agreement with experiment, the simulated anisotropy of hopping following a 0.7-THz pump exhibits a long-lived relaxation at 300 K, but not at 600 K (Figure 4b). The simulation reproduces the most important feature of experimental TKE: a picosecond-timescale anisotropy of hopping at 300 K that disappears by 600 K. The experiment and simulation are consistent with the expected change of transport mechanism: correlated at 300 K changing to a true random walk by 600 K[7]. We conclude that the non-oscillatory picosecond-timescale relaxation of the terahertz Kerr effect indeed probes the decay of orientational memory in ionic hopping at the level of individual ionic hops.

## Attempt Frequencies for Ion Hopping

Finally, we use the TKE measurements and simulations of K+ $\beta$"-alumina to identify the vibrational origin of ionic hops: the attempt frequency $v_0$. To cause an anisotropy in hopping, a pump field must excite the vibrational modes $Q_i$ that couple directly to hopping, and trigger hopping *aligned* with the applied field. This must happen before thermalization, or else pump-driven heating enhances hopping rates in all directions isotropically. Having established the correspondence of the experimental TKE and simulated anisotropy of ionic hopping, we use further molecular dynamics simulations as *in silico* TKE at pump frequencies presently inaccessible to us experimentally. We simulate the anisotropy of hopping in K+ $\beta$"-alumina following a 2.5-THz pulse that overlaps with the strong infrared-active vibration of the K+ ions in the conduction plane at 3.0 THz (Figure 4c, orange). Despite the material absorbing nearly the same amount of energy from simulated 0.7-THz and 2.5-THz pulses (Figure S6), and despite the 3.0-THz vibration being coherently driven by the 2.5-THz pulse (Figure S7), the 0.7-THz pulse creates a ≈20x stronger anisotropy of hopping than the 2.5-

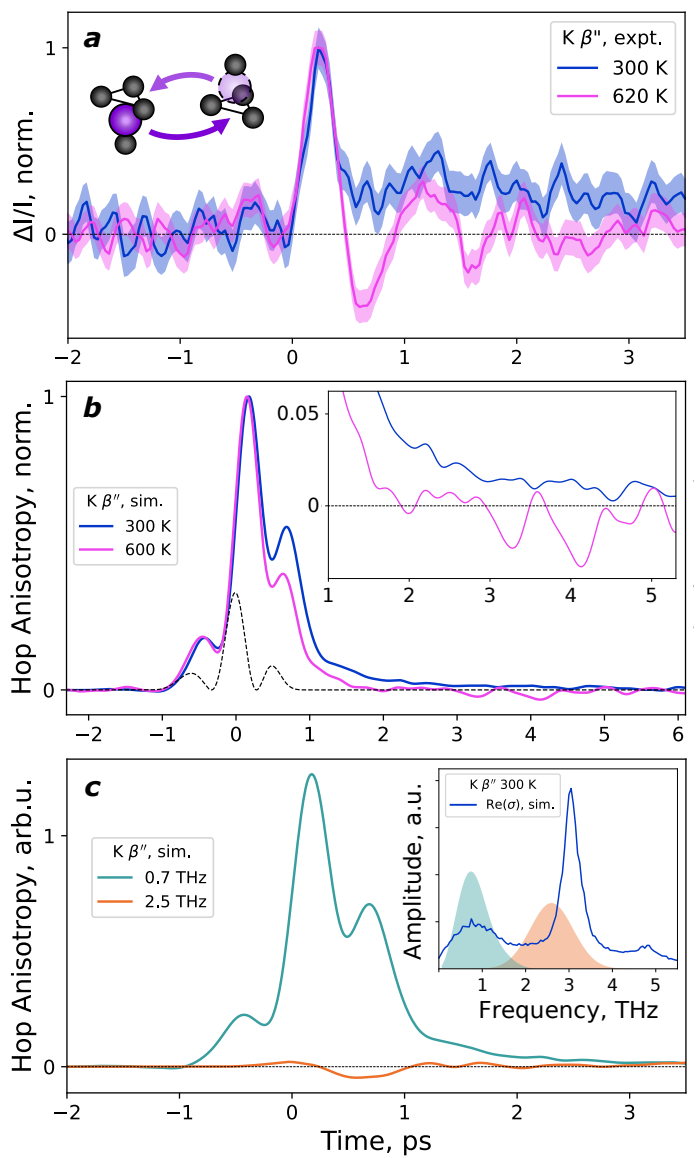

**Figure 4 | Terahertz Kerr effect in K+ $\beta$"-alumina**. (**a**) Normalized time-domain transient birefringence in thin polycrystalline K+ $\beta$"-alumina at 300 K (blue) and 620 K (pink). The shaded regions are ±1 s.e. of the mean. Inset: in $\beta$"-aluminas mobile ions (purple) hop (arrows) between equivalent four-coordinate sites. (**b**) Normalized simulated anisotropy of hopping directions in polycrystalline K+ $\beta$"-alumina under applied electric field centered at 0.7 THz mimicking the experimental pump, 300 K (blue) and 600 K (pink). Black dashed line: square of the simulated pump electric field. Inset zooms in on the post-pump times: simulated hopping anisotropy at 300 K is nonzero for ≥5 ps. (**c**) Simulated anisotropy of hopping in polycrystalline K+ $\beta$"-alumina at 300 K under applied electric fields centered at 0.7 THz (teal), and 2.5 THz (orange). Inset: Simulated in-plane optical K+ conductivity of K+ $\beta$"-alumina in the terahertz region (blue). Peaks at ≈1, 3, and ≈5 THz are in agreement with literature[50]. Shaded: spectra of simulated pumps with center frequencies 0.7 THz (teal) mimicking the experimental pump (**ab**), and 2.5 THz (orange).

THz pulse (Figure 4c). This suggests that the pulse at ≈1 THz couples directly to hops, whereas the 2.5-THz pulse heats the material isotropically. We conclude that the attempt frequency in K$^+$ $\beta$"-alumina is ≈1 THz. The excitation of this vibration is evident in the experimental TKE response (Figure 4a, Extended Data Figure 7b). The attempt frequencies in $\beta$-aluminas can be similarly identified once the distinct dynamics of bound defect clusters and non-equivalent lattice sites are dis-aggregated (Supplementary Note 6). For example, in Na $\beta$-alumina the simulated attempt frequencies are ≈1.2 THz for the dominant ion-pair hopping mechanism, and ≈2.3 THz for rare single-ion hopping. The latter corresponds to the known vibration[37] commonly taken as the attempt frequency[41] and verified here with THz transmission (2.1 THz, Figure S5, and Supplementary Note 6). Instead, the 1.2-THz vibration, highlighted by neutron scattering[36], drives most hopping at 300K, despite constituting a minor part of the vibrational density of states for the mobile ion.

## Conclusions

In summary, we have used impulsive near-resonant terahertz excitation to trigger ionic hopping in solid-state ionic conductors. Picosecond transient birefringence arises from correlations in the hopping of mobile ions. Therefore we establish TKE, a nonlinear optical measurement, as a direct probe of path dependence in ionic transport[7,8]. TKE highlights both the vibrational origination of ionic conduction, and the the slow loss of memory during diffusion. Ionic conductivity and its activation reach their low-frequency limits only at timescales sufficiently long to scramble all memory within the system. The transport of ions can be characterized by a random walk only at timescales longer than the persistence of correlations - but such a phenomenological random walk may not correspond to a true atomistic one. In other words, macroscopic measurements cannot be interpreted in terms of atomistic quantities without accounting for such correlations. By distinguishing rapid ion hops from the persistent correlations connecting them, our work demonstrates the correspondence between thermodynamics and information for thermally activated mass transport.

In addition to probing the mechanisms of ionic conduction, the correlation effects we highlight are of importance for transport under strong driving forces, at short timescales, and in confined dimensions[9,10], such as in switching applications[3]. This study provides a framework for the use of nonlinear optical techniques to probe the atomistic mechanisms of non-equilibrium transport phenomena, which will be highly valuable in the development of energy technologies such as solid state batteries, and in the related fields of nanofluidics, engineering phase transformations, and neuromorphic computing.

## Acknowledgements

This work was supported by the U.S. Department of Energy, Office of Basic Energy Sciences, Division of Materials Sciences and Engineering (contract DE-AC02-76SF00515). MSI and JAD gratefully acknowledge the EPSRC Programme Grant "Enabling next generation lithium batteries" (EP/M009521/1). JAD gratefully acknowledges the EPSRC (EP/V013130/1), Research England (Newcastle University Centre for Energy QR Strategic Priorities Fund) and Newcastle University (Newcastle Academic Track (NUAcT) Fellowship) for funding. We are indebted to Prof. Osamu Kamishima (Setsunan University) for sharing single-crystalline samples of Na $\beta$-alumina with us.

## Author Contributions

The probing of ionic transport with nonlinear techniques was proposed by AML, MT, SWT, and ADP. Simulations incorporating the terahertz pump were proposed by AML, JAD, and MSI. ADP and MCH performed the terahertz Kerr effect and transmission measurements. ADP, SWT, and MT performed the optical Kerr effect measurements. ADP analyzed the experimental data. ADP performed and analyzed the molecular dynamics simulations with advice from JAD and MSI. AML advised and supervised the work. All authors contributed to the writing of the manuscript.

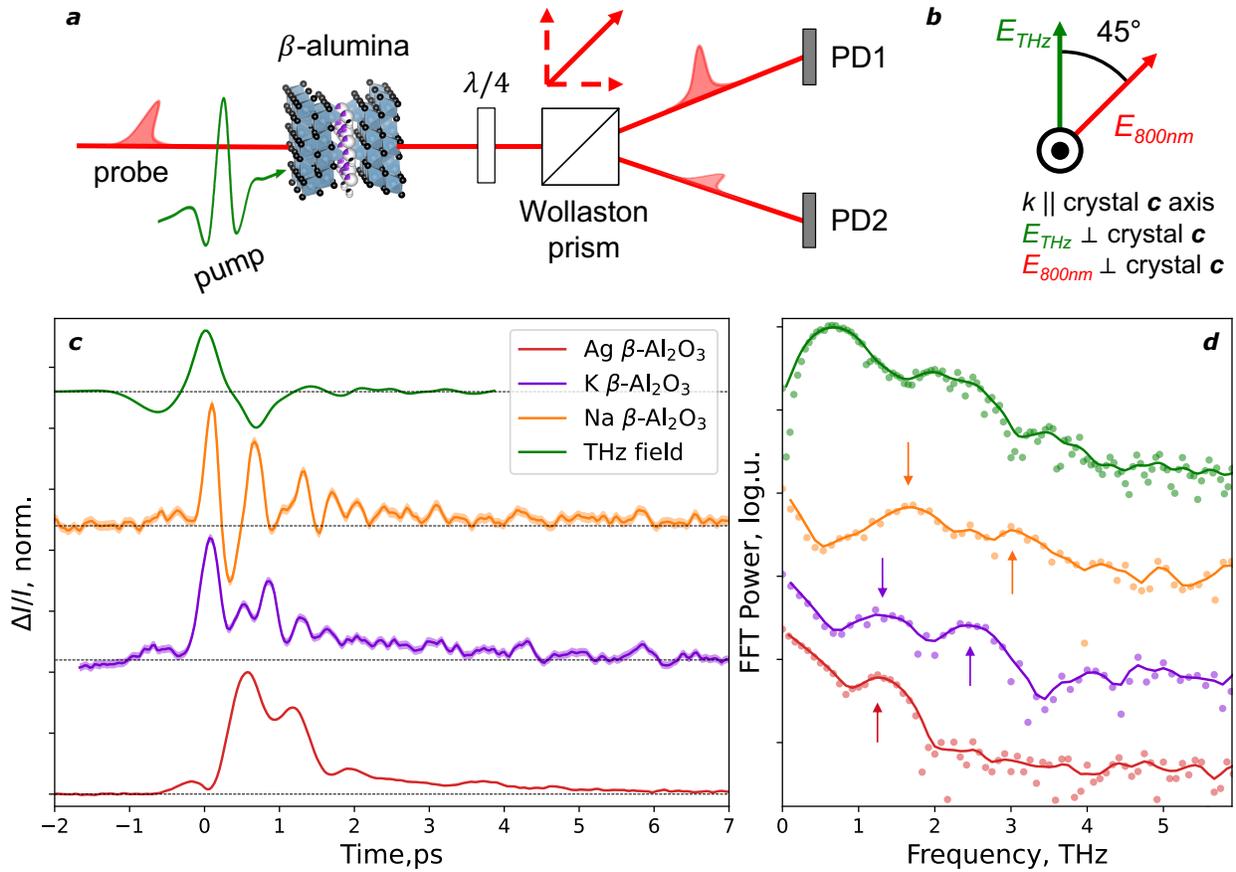

**Extended Data Figure 1 | Terahertz Kerr effect (TKE) in $\beta$-aluminas at elevated temperature.** (**a**) Schematic of the transient birefringence experiment: terahertz pump (green) perturbs mobile ions (purple) in the ionic conductor. The 800 nm probe pulse (pink) delayed by a time $\Delta t$ is split with a quarter wave plate and a Wollaston prism. The polarization rotation of the probe pulse is detected with a pair of Si photodiodes in a balanced detection scheme. (**b**) Polarizations of the pump and probe pulses relative to each other and the $\beta$-alumina crystals. (**c**) Time-domain traces of transient birefringence in thin single crystals of $\beta$-aluminas at 620 K: $Na^+$ (orange), $K^+$ (purple), $Ag^+$ (dark red), and electro-optic sampling trace of the pump field (green). The shaded regions correspond to ±1 s.e. of the mean signal at each time delay. (**d**) Fourier transform power spectra (points) of the signals in (**c**), and smoothed with a Gaussian filter of 0.1 THz st. dev. (lines), with peaks highlighted by arrows.

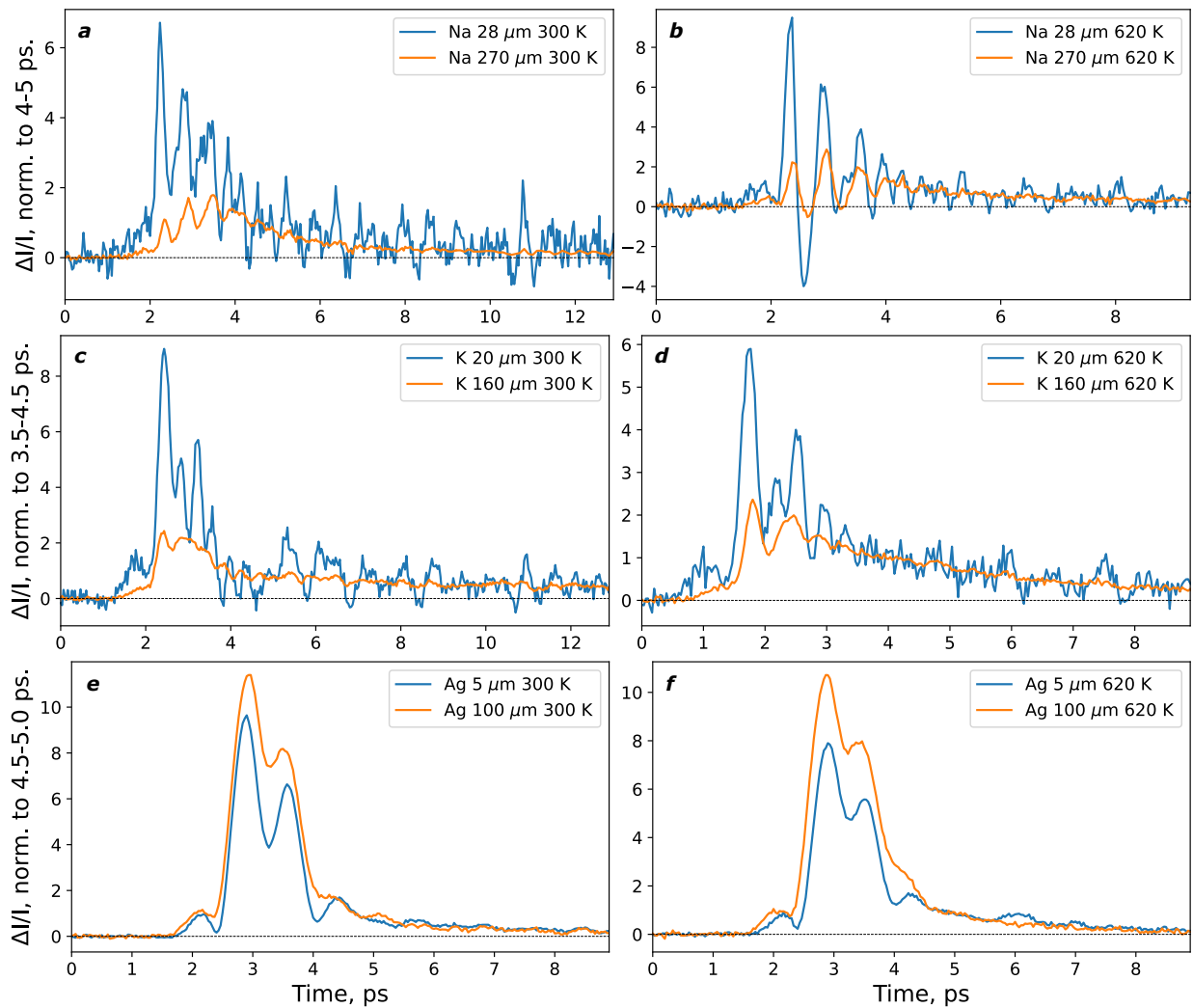

**Extended Data Figure 2 | TKE in $\beta$-aluminas with varying sample thickness.** TKE signals for thick (≥100 μm) and thin (≤30 μm) crystals of Na (**ab**), K (**cd**), and Ag (**ef**) $\beta$-alumina, normalized to their values at a pump-probe delay time when short-time oscillatory signal components have decayed. For all materials at 300 K (**ace**) and 620 K (**bdf**) the longer-time relaxation of the signals overlap: their kinetics are independent of sample thickness. Since the TKE signal is phase accumulated over the sample thickness, thicker samples yield lower-noise signals at long time delays.

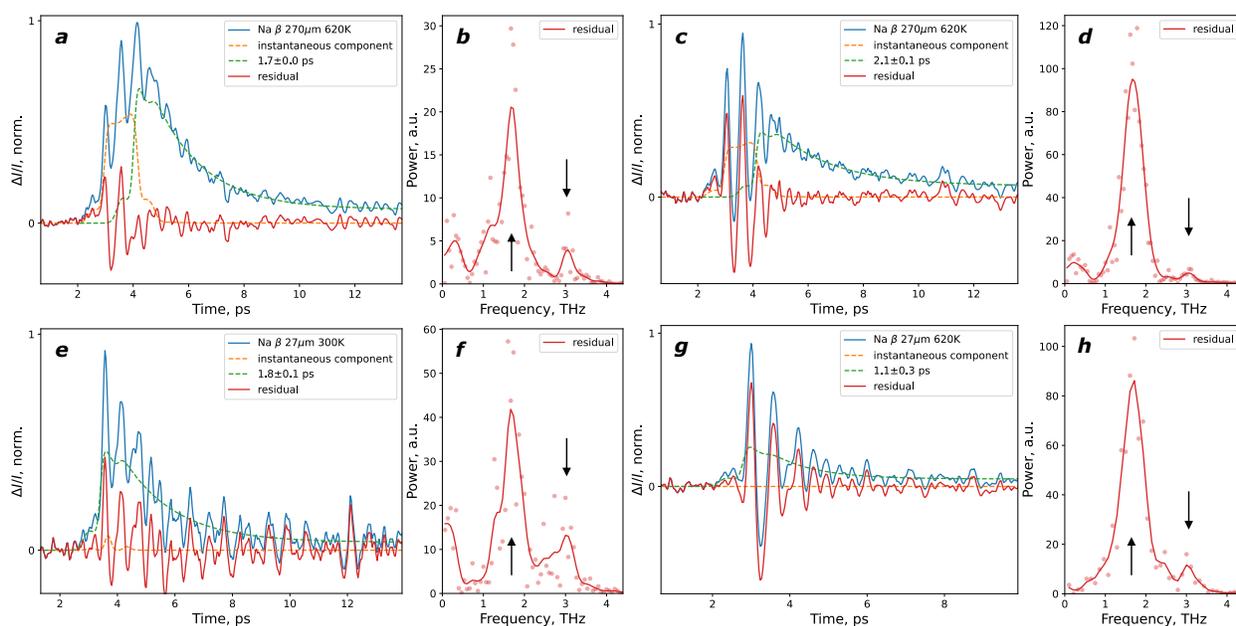

**Extended Data Figure 3 | Fitting TKE signals of Na $\beta$-alumina**. Each signal (blue in **aceg**) was measured in a purged atmosphere and modeled as the sum of an instantaneous component (orange) and a single-exponential relaxation (green) with a long-time constant (Supplementary Notes 1 and 2). For the thin sample (**e-h**), the strength of the instantaneous component was small. The strong vibrational component in the thick-sample signal (a-d) precluded unambiguous fitting at short times, so the exponential component was fit to the long-time part of the signal. The residuals (dark red) are oscillatory. Their Fourier transforms (**bdfh**), plotted with 0.1-THz Gaussian-filter smoothing, show main frequency components at 1.8 and 3.0 THz in agreement with non-resonant OKE (3.0 THz, Extended Data Figure 6), literature, and simulation (1.7 THz, Figure S7).

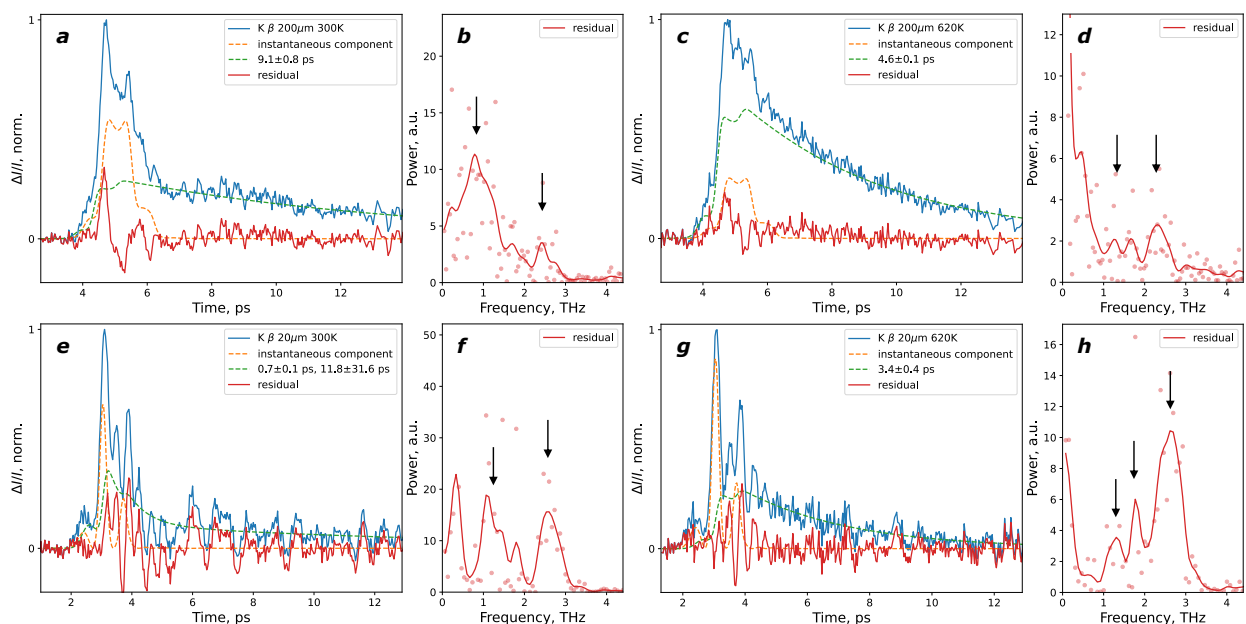

**Extended Data Figure 4 | Fitting TKE signals of K *β*-alumina**. Each signal (blue in **aceg**) was measured in a purged atmosphere and modeled as the sum of an instantaneous component (orange) and a single-exponential relaxation (green) with a long-time constant (Supplementary Notes 1 and 2). For the thin sample (**e-h**), the instantaneous component was non-negligible, and at 300 K a weakly determined second exponential was identifiable. The residuals (dark red) are oscillatory and most interpretable in the signals from the thin sample (**e-h**). Their Fourier transforms (**bdfh**), plotted with 0.1-THz Gaussian-filter smoothing, show main frequency components at ≈1.2, ≈2.0, and ≈2.7 THz in agreement with non-resonant OKE (2.0 THz, Extended Data Figure 6), literature, and simulation (1.3 THz and 1.9 THz, Figure S7).

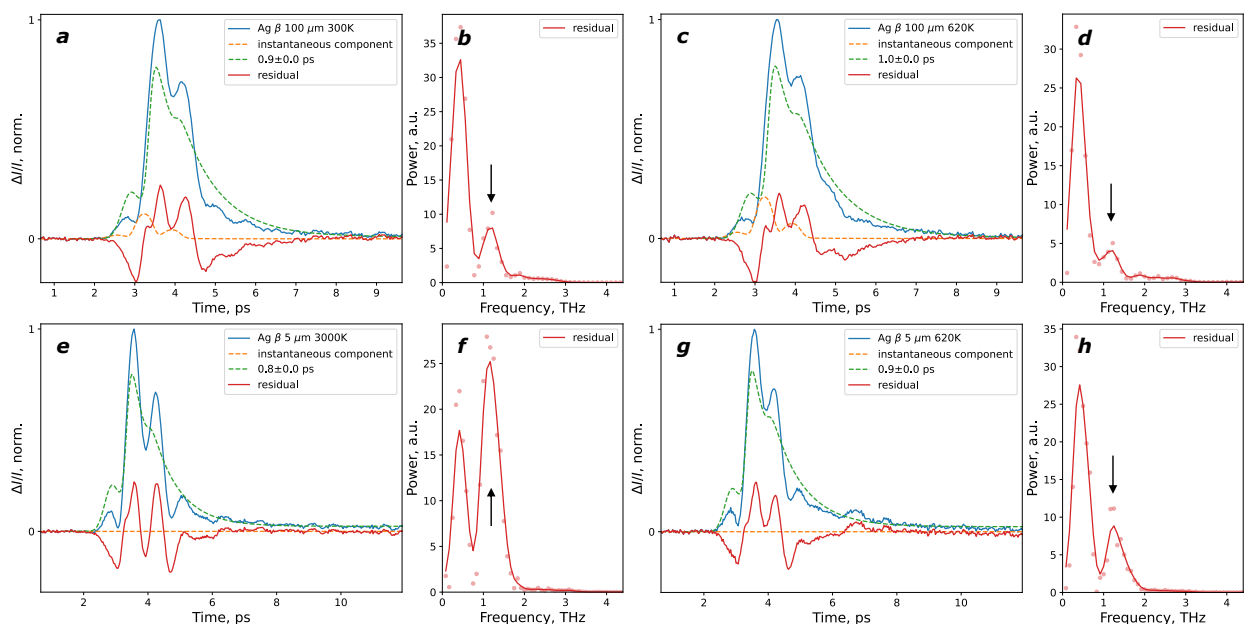

**Extended Data Figure 5 | Fitting TKE signals of Ag $\beta$-alumina**. Each signal (blue in **aceg**) was measured in a purged atmosphere and modeled as the sum of an instantaneous component (orange) and a single-exponential relaxation (green) (Supplementary Notes 1 and 2). For the thin sample (**e-h**), the instantaneous component was negligible. The Fourier transforms (**bdfh**) of residuals (dark red), plotted with 0.1-THz Gaussian-filter smoothing, show a frequency component at ≈1.2 THz in agreement with literature (infrared-active mode at 1.0-1.2 THz) and simulation (1.0 THz, Figure S7). The signal is very similar between the thick and thin samples (Extended Data Figure 2), consistent with it arising from a thin layer of the material and consistent with strong absorption (Supplementary Note 3).

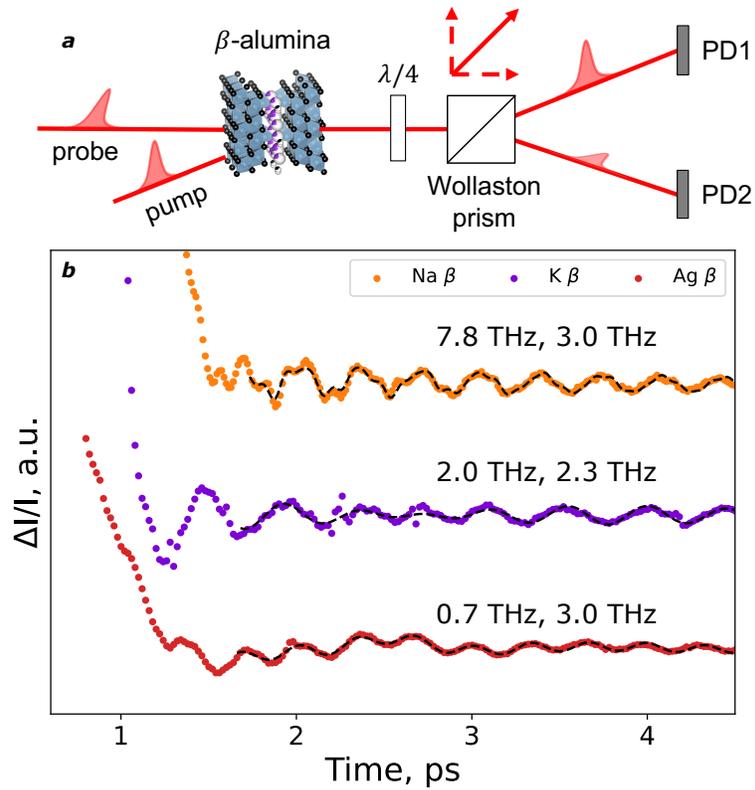

**Extended Data Figure 6 | Non-resonant optical Kerr effect in $\beta$-aluminas.** Schematic of the experimental configuration (**a**) and signals from thick Na (orange), K (purple), and Ag (dark red) $\beta$-alumina crystals (**b**) at positive time delays following a strong coherent artifact at zero time delay. The frequency components are extracted from linear prediction fitting results (Methods) shown as black dashed lines.

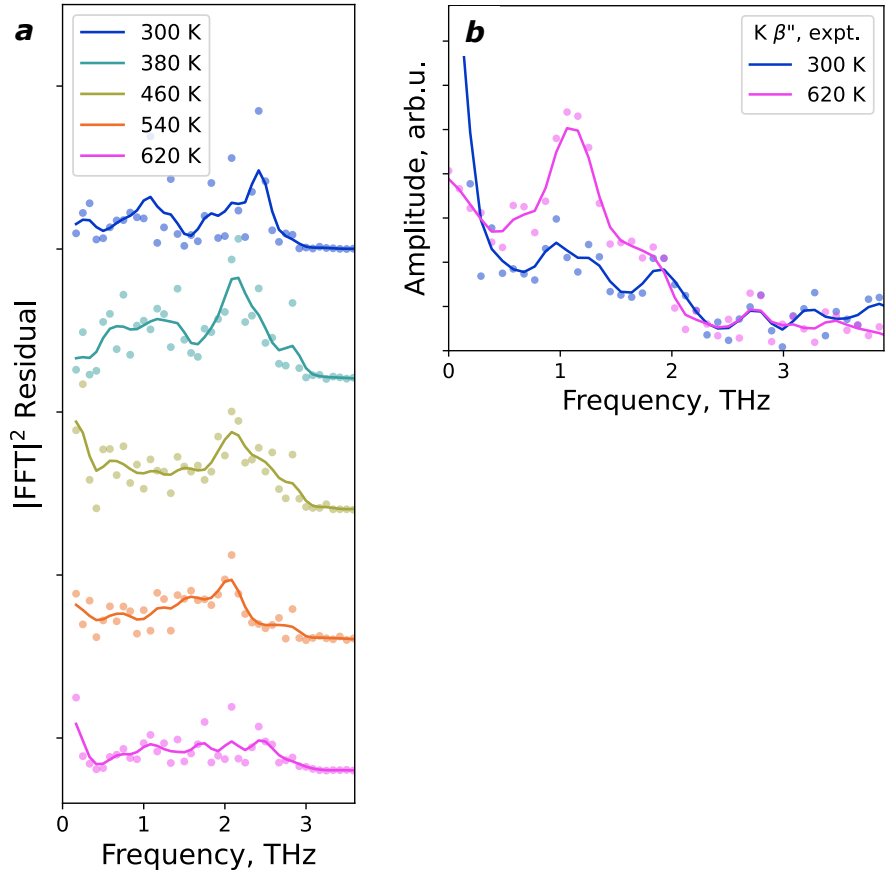

**Extended Data Figure 7 | Fourier Transforms of TKE Signals in K⁺ $\beta$- and $\beta$"-alumina**. (**a**) Fourier transforms of oscillatory residuals in TKE signals of thick crystals of K $\beta$-alumina measured in ambient atmosphere (Figure 3a). (**b**) Fourier transforms of the TKE signals of thin polycrystalline K⁺ $\beta$"-alumina (Figure 4a).

## References

1. Famprikis, T., Canepa, P., Dawson, J. A., Islam, M. S. & Masquelier, C. Fundamentals of inorganic solid-state electrolytes for batteries. *Nat. Mater.* **18**, 1278 (2019).
2. Ohno, S. *et al.* Materials design of ionic conductors for solid state batteries. *Prog. Energy* **2**, 022001 (2020).
3. Sood, A. *et al.* Electrochemical ion insertion from the atomic to the device scale. *Nat. Rev. Mater.* **6**, 847–867 (2021).
4. He, X., Zhu, Y. & Mo, Y. Origin of fast ion diffusion in super-ionic conductors. *Nat. Commun.* **8**, 15893 (2017).
5. Bachman, J. C. *et al.* Inorganic Solid-State Electrolytes for Lithium Batteries: Mechanisms and Properties Governing Ion Conduction. *Chem. Rev.* **116**, 140–162 (2016).
6. Funke, K., Cramer, C. & Wilmer, D. Concept of mismatch and relaxation for self-diffusion and conduction in ionic materials with disordered structures. in *Diffusion in Condensed Matter* 857–893 (Springer-Verlag, 2005). doi:10.1007/3-540-30970-5_21.
7. Poletayev, A. D., Dawson, J. A., Islam, M. S. & Lindenberg, A. M. Defect-Driven Anomalous Transport in Fast-Ion Conducting Solid Electrolytes. *https://arxiv.org/abs/2105.08761* (2021).
8. Song, S. *et al.* Transport dynamics of complex fluids. *Proc. Natl. Acad. Sci.* **116**, 12733–12742 (2019).
9. Kavokine, N., Netz, R. R. & Bocquet, L. Fluids at the Nanoscale: From Continuum to Subcontinuum Transport. *Annu. Rev. Fluid Mech.* **53**, 377–410 (2021).
10. Agarwal, R. K., Yun, K. Y. & Balakrishnan, R. Beyond Navier-Stokes: Burnett equations for flows in the continuum-transition regime. *Phys. Fluids* **13**, 3061–3085 (2001).
11. Gao, Y. *et al.* Classical and Emerging Characterization Techniques for Investigation of Ion Transport Mechanisms in Crystalline Fast Ionic Conductors. *Chem. Rev.* **120**, 5954–6008 (2020).
12. Muy, S., Schlem, R., Shao-Horn, Y. & Zeier, W. G. Phonon–Ion Interactions: Designing Ion Mobility Based on Lattice Dynamics. *Adv. Energy Mater.* **11**, 2002787 (2021).
13. Franco, A. A. *et al.* Boosting Rechargeable Batteries R&D by Multiscale Modeling: Myth or Reality? *Chem. Rev.* **119**, 4569–4627 (2019).
14. Maier, J. *Physical Chemistry of Ionic Materials: Ions and Electrons in Solids*. *John Wiley & Sons, Ltd* vol. 1 (John Wiley & Sons, Ltd, 2004).
15. Murch, G. E. The Haven Ratio In Fast Ionic Conductors. *Solid State Ionics* **7**, 177–198 (1982).
16. Vargas-Barbosa, N. M. & Roling, B. Dynamic Ion Correlations in Solid and Liquid Electrolytes: How Do They Affect Charge and Mass Transport? *ChemElectroChem* **7**, 367–385 (2020).
17. Cover, T. M. & Thomas, J. A. *Elements of Information Theory*. (Wiley, 2005). doi:10.1002/047174882X.
18. Coffey, W. T. & Kalmykov, Y. P. *The Langevin Equation*. *World Scientific Series in Contemporary Chemical Physics* vol. 28 (2017).
19. Klafter, J. & Sokolov, I. M. *First Steps in Random Walks*. (Oxford University Press, 2011). doi:10.1093/acprof:oso/9780199234868.001.0001.
20. Habasaki, J., Leon, C. & Ngai, K. L. Electrical response of ionic conductors. in *Dynamics of Glassy, Crystalline and Liquid Ionic Conductors* 89–250 (Springer Verlag, 2017).

doi:10.1007/978-3-319-42391-3_4.
21. Funke, K. & Banhatti, R. D. Conductivity spectroscopy covering 17 decades on the frequency scale. *Solid State Ionics* **176**, 1971–1978 (2005).
22. Kamishima, O. *et al.* Temperature dependence of low-lying phonon dephasing by ultrafast spectroscopy (optical Kerr effect) in Ag β-alumina and Tl β-alumina. *J. Phys. Condens. Matter* **19**, 456215 (2007).
23. Parrondo, J. M. R., Horowitz, J. M. & Sagawa, T. Thermodynamics of information. *Nat. Phys.* **11**, 131–139 (2015).
24. Ciliberto, S. Experiments in Stochastic Thermodynamics: Short History and Perspectives. *Phys. Rev. X* **7**, 16–21 (2017).
25. Krauskopf, T. *et al.* Comparing the Descriptors for Investigating the Influence of Lattice Dynamics on Ionic Transport Using the Superionic Conductor Na 3 PS 4-x Se x. *J. Am. Chem. Soc.* **140**, 14464–14473 (2018).
26. Muy, S. *et al.* Tuning mobility and stability of lithium ion conductors based on lattice dynamics. *Energy Environ. Sci.* **11**, 850–859 (2018).
27. Elgabarty, H. *et al.* Energy transfer within the hydrogen bonding network of water following resonant terahertz excitation. *Sci. Adv.* **6**, 1–15 (2020).
28. Zalden, P. *et al.* Molecular polarizability anisotropy of liquid water revealed by terahertz-induced transient orientation. *Nat. Commun.* **9**, 1–7 (2018).
29. Hoffmann, M. C., Brandt, N. C., Hwang, H. Y., Yeh, K.-L. & Nelson, K. A. Terahertz Kerr effect. *Appl. Phys. Lett.* **95**, 1–4 (2009).
30. de la Torre, A. *et al.* Colloquium : Nonthermal pathways to ultrafast control in quantum materials. *Rev. Mod. Phys.* **93**, (2021).
31. Mankowsky, R., Först, M. & Cavalleri, A. Non-equilibrium control of complex solids by nonlinear phononics. *Reports Prog. Phys.* **79**, 064503 (2016).
32. Hebling, J., Yeh, K.-L., Hoffmann, M. C. & Nelson, K. A. High-power THz generation, THz nonlinear optics, and THz nonlinear spectroscopy. *IEEE J. Sel. Top. Quantum Electron.* **14**, 345–353 (2008).
33. Hoffmann, M. C. & Fülöp, J. A. Intense ultrashort terahertz pulses: Generation and applications. *J. Phys. D. Appl. Phys.* **44**, (2011).
34. Yan, Y. X., Gamble, E. B. & Nelson, K. A. Impulsive stimulated scattering: General importance in femtosecond laser pulse interactions with matter, and spectroscopic applications. *J. Chem. Phys.* **83**, 5391–5399 (1985).
35. Merlin, R. Generating Coherent THz Phonons with Light Pulses. *Solid State Commun.* **102**, 207–220 (1997).
36. McWhan, D. B., Shapiro, S. M., Remeika, J. P. & Shirane, G. Neutron-scattering studies on beta-alumina. *J. Phys. C Solid State Phys.* **8**, L487 (1975).
37. Lucazeau, G. Infrared, Raman and neutron scattering studies of β- and β″-alumina: a static and dynamical structure analysis. *Solid State Ionics* **8**, 1–25 (1983).
38. Sajadi, M., Wolf, M. & Kampfrath, T. Transient birefringence of liquids induced by terahertz electric-field torque on permanent molecular dipoles. *Nat. Commun.* **8**, 1–8 (2017).
39. Allodi, M. A., Finneran, I. A. & Blake, G. A. Nonlinear terahertz coherent excitation of vibrational modes of liquids. *J. Chem. Phys.* **143**, (2015).
40. Zhu, H. *et al.* Screening in crystalline liquids protects energetic carriers in hybrid perovskites. *Science (80-. ).* **353**, 1409–1413 (2016).

41. Minami, Y. *et al.* Macroscopic Ionic Flow in a Superionic Conductor Na+ β-Alumina Driven by Single-Cycle Terahertz Pulses. *Phys. Rev. Lett.* **124**, 147401 (2020).
42. Neugebauer, M. J. *et al.* Comparison of coherent phonon generation by electronic and ionic Raman scattering in LaAlO$_3$. *Phys. Rev. Res.* **3**, 13126 (2021).
43. Först, M. *et al.* Nonlinear phononics as an ultrafast route to lattice control. *Nat. Phys.* **7**, 854–856 (2011).
44. Fleischer, S., Zhou, Y., Field, R. W. & Nelson, K. A. Molecular orientation and alignment by intense single-cycle THz pulses. *Phys. Rev. Lett.* **107**, 1–5 (2011).
45. Sajadi, M., Wolf, M. & Kampfrath, T. Terahertz-field-induced optical birefringence in common window and substrate materials. *Opt. Express* **23**, 28985 (2015).
46. Maehrlein, S. F. *et al.* Decoding ultrafast polarization responses in lead halide perovskites by the two-dimensional optical Kerr effect. *Proc. Natl. Acad. Sci.* **118**, e2022268118 (2021).
47. Mishra, P. K., Vendrell, O. & Santra, R. Ultrafast Energy Transfer from Solvent to Solute Induced by Subpicosecond Highly Intense THz Pulses. *J. Phys. Chem. B* **119**, 8080–8086 (2015).
48. Mishra, P. K., Bettaque, V., Vendrell, O., Santra, R. & Welsch, R. Prospects of Using High-Intensity THz Pulses to Induce Ultrafast Temperature-Jumps in Liquid Water. *J. Phys. Chem. A* **122**, 5211–5222 (2018).
49. Whittingham, M. S. & Huggins, R. A. Electrochemical preparation and characterization of alkali metal tungsten bronzes, MxWO3. in *Proceedings of the 5th materials research symposium sponsored by the Institute for Materials Research, National Bureau of Standards* (eds. Roth, R. S. & Schneider S.J. Jr) 51–62 (1972).
50. Hayes, W., Hopper, G. F. & Pratt, F. L. Ionic conductivity of potassium β" alumina in the very far infrared. *J. Phys. C Solid State Phys.* **15**, L675–L680 (1982).

# The Persistence of Memory in Ionic Conduction Probed by Nonlinear Optics: Supplementary Information

Methods
Supplementary Note 1: Terahertz Kerr Effect (TKE)
Supplementary Note 2: TKE Control Measurements
Supplementary Note 3: Optical Conductivity from Terahertz Transmission
Supplementary Note 4: Terahertz Pumps in Molecular Dynamics
Supplementary Note 5: Logical and Thermodynamic Reversibility
Supplementary Note 6: Attempt Frequencies in Na $\beta$-alumina

Figures S1-S11
Supplementary References 1-81

## Methods

### Sample Preparation
Single crystals of Na $\beta$-alumina were graciously shared with us by Prof. Osamu Kamishima. They were ion-exchanged to Ag and K in molten nitrates until no mass change was detectable, at least 3 days. For Ag, the process had to be repeated. For K, mixed nitrate compositions were used first to avoid mechanical damage from the thermodynamically favorable ion exchange. The *c* lattice constants for the Na, Ag, and K $\beta$-alumina samples were 22.53, 22.49, and 22.73 Å, respectively. Polycrystalline K $\beta$"-alumina was purchased from Ionotec Ltd as a pellet. To produce thin samples (≈5-30 μm), single crystals and the polycrystalline material were hand-polished using a T-tool, and then dried. Double-side polished sapphire (0001) was purchased from MTI Corp.

### Terahertz Kerr Effect (TKE)
The output of a Ti:sapphire oscillator (Coherent Micra) is amplified (Coherent Spitfire) to ≈4.2 mJ at 1 kHz, and pulse width optimized to maximize the peak terahertz field at the sample, ≈150 fs full width at half maximum. 99% of the output is used to generate terahertz pulses (≈6 μJ) via optical rectification in lithium niobate using the tilted pulse front method[1,2]. The terahertz pulse is focused on the sample using a pair of off-axis parabolic (OAP) mirrors. Peak field amplitudes at the sample position were ≈700 kV/cm in ambient atmosphere, and ≈600 kV/cm in purged (≤0.1% RH) atmosphere. 1% of the amplifier output is used for the probe pulse, polarized at 45 degrees from the pump pulse, overlapped with the pump at the sample position, and passed through a quarter-wave plate and a Wollaston prism. The birefringence of the transmitted probe is measured with two photodiodes in a balanced detection scheme. The terahertz waveform at the sample position is measured with a free-standing uniaxially poled mixture of an electro-optic dye and amorphous polycarbonate polymer[3,4], and the peak field strength measured by electro-optic sampling in GaP(110). The time delay between pump and probe pulses is varied with a mechanical delay stage. For temperature control and measurements at elevated temperature, a transmission-mode heating stage (Linkam) was used without windows to eliminate their contributions to transient birefringence. The transient birefringence of ambient and purged air (Supplementary Note 2) was measured by aligning to a thin sample and removing the sample. The majority of experimental data is adapted from ADP's doctoral dissertation[5].

## Terahertz Transmission

Thin (≤ 30 μm) samples are mounted on thin metallic pinholes of diameter 1 mm, fully covering the pinhole. The terahertz pulse transmitted through the sample is focused with a second pair of OAP mirrors and sampled with the free-standing film of an electro-optic dye and amorphous polycarbonate polymer[3,4]. A time delay sweep of an empty pinhole is measured following every time delay sweep measuring the sample. Optical conductivity was fit to the time-domain spectra assuming a slab geometry[6,7] using a nonlinear fitting procedure initialized with $ñ=2+0.5j$ as the complex refractive index. This yielded stable fitting without the single-pass assumption.

## Optical Kerr Effect (OKE)

The output of a Ti:sapphire oscillator (Coherent Micra) was amplified (Coherent RegA) at 100 kHz to ≈1 μJ and compressed to ≈50 fs FWHM (the width of the coherent artifact in OKE). The pump and probe pulses were overlapped on the sample so that probe $E \perp c$ for single crystals, and pump ≈15 degrees off. The transmitted probe was detected using a pair of Si photodetectors in a balanced detection scheme. Frequencies of oscillatory components in the OKE signals were fit using linear prediction fitting[8] optimizing for least-squares error over the start and end points, which serve as hyperparameters.

## Steady-State Molecular Dynamics (MD)

Steady-state classical MD simulations were carried out using LAMMPS[9] using Buckingham pairwise potentials with Coulombic interactions as described previously[10]. The vibrational density of states was calculated from the velocity autocorrelation functions for the mobile ions, and the optical conductivity from the velocity autocorrelation function for the center of mass of the mobile ions[11].

## Molecular Dynamics with Terahertz Pumping

An impulsive electrical field was included in the molecular dynamics simulations to mimic the experimental terahertz pulse[12–15]. Peak field $E$ = 300 kV/cm polarized parallel to the conduction plane was used to account for ≈50% front-surface reflection. Charge-compensating defects were placed in the scaled-up simulations using the same procedure as described previously[10]. To sample hopping events, 800 randomized iterations per temperature point, with 4896 mobile ions each, were used for β-aluminas. For K β"-alumina, the simulations contained 5940 mobile ions each. Each iteration started with an anneal to 1000 K to randomize starting positions of the mobile ions, and a short equilibration at the simulation temperature in the NVT ensemble. The electric field was applied in the NVE ensemble, and the simulations propagated for 15 ps, enough for mobile ions and the host lattice to come to the same temperature. For K β"-alumina, polycrystal orientations to match the experimentally available sample were averaged by changing the polarization of the simulated field over 12 angles in increments of 30 degrees. No significant dependence on the pump polarization within the conduction plane was observed in simulations of β-aluminas. The hopping statistics were extracted from all simulations and analyzed as described previously[10]. The direction of each hopping event was taken as the vector connecting the start and end crystallographic sites of the hop. The anisotropy of hopping was calculated by weighing the hopping directions by $(\cos\theta)^2 - \langle(\cos\theta)^2\rangle$, with $\theta$ the angle between the simulated pump field and the hop direction, and $\langle(\cos\theta)^2\rangle = 0.5$ in two dimensions.

## Supplementary Note 1: Terahertz Kerr Effect

The 2D terahertz Kerr effect measurement is a $\chi^{(3)}$ measurement that relies on the third-order response function[16,17]:

$$R^{(3)}_{ijkl}(t'', t') = \left(\frac{i}{\hbar}\right)^2 tr\langle \alpha_{ij}(t''+t')[\mu_k(t')[\mu_l(0), \rho]]\rangle$$

Here, the transition dipole moment operator $\mu$ and polarizability operator $\alpha$ act on the equilibrium density matrix $\rho$ describing the thermal population of vibrational states. The pump pulses arrive at times $0$ and $t'$. The 800nm probe pulse at $t''+t'$ is assumed to be short. Because we use one pump pulse, our experiment represents a line cut through the ($t'$, $t''$) space corresponding to $t'=0$. The third-order polarization in the sample is given by the integral over pathways:

$$P_i(t) = \int\int dt'dt'' E_{probe,j}(t) E_{THz,k}(t-t'') E_{THz,l}(t-t'-t'')$$

This formalism includes all possible excitation pathways via either bound vibrational states, virtual states (Raman), or "thermal bath" states[16,17]. The generated field is the time derivative of the overall ionic and electronic polarization[18–20]:

$$E_{probe}(z,t) - E(z,t) \propto \frac{\partial}{\partial t}\left(P(z, t+t'') E_{probe}(z,t)\right)$$

Models of coherently excited vibrational responses utilize an oscillatory form of the generated polarization corresponding to a bound vibrational state, including possible anharmonic potential energy surfaces[20,21], anharmonic couplings to other vibrations[16,22], or couplings to a thermal bath[15,23]. For such an oscillating polarization, the detected change in the probe field at any time is proportional to the velocity of the oscillator. However, this formalism does not impose a coherence requirement or a functional form on the trajectory of the charged species interacting with the pump and probe fields. In principle, this treatment should be valid for Debye relaxation[15] or an ensemble of coupled double-well oscillators coupled to a thermal bath and stochastically switching their positions. Here, the ionic polarization $P$ arises from coherent displacements of vibrational modes or bound dipoles $Q_i$, plus an additional incoherent component representing hopping rates $H$. The detected third-order signal is proportional to the transient anisotropy of the hopping rate, measured as a function of the pump-probe time delay. The pump-probe time delay is $t''$ in the above formalism. The different regimes of sample responses are discussed in the main text and Figure 1 for short time delays $t''=t_1$, when hopping is correlated, and longer time delays $t''=t_2$, when hopping reaches the random walk limit.

The probe signal is summed across the thickness $L$ of the sample. For mismatched group velocities between the pump and probe pulses, and for $t'' < L/\Delta v_g$, where $\Delta v_g$ is the mismatch of the group velocities of the pump and probe pulses[20,24–26], the probe overtakes the pump within the sample, and samples the instantaneous electronic polarization in addition to the ionic responses. For a terahertz pump and 800 nm probe in a sample mostly composed of alumina, refractive indices are taken as $n_{probe} \approx 1.76$, and $n_{pump} \approx 3.08$[24,27], i.e. the probe propagates faster than the pump inside the sample. Ignoring the absorption of the pump pulse, and for the probe pulse assumed short, the strength of this instantaneous electronic-origin signal is proportional to the thickness of the sample convoluted with the square of the pump field and the nonlinear refractive index of the sample[24]. In fitting the TKE signals of β-aluminas and K β"-alumina, this component is referred to as the "instantaneous component".

## Supplementary Note 2: TKE Control Measurements

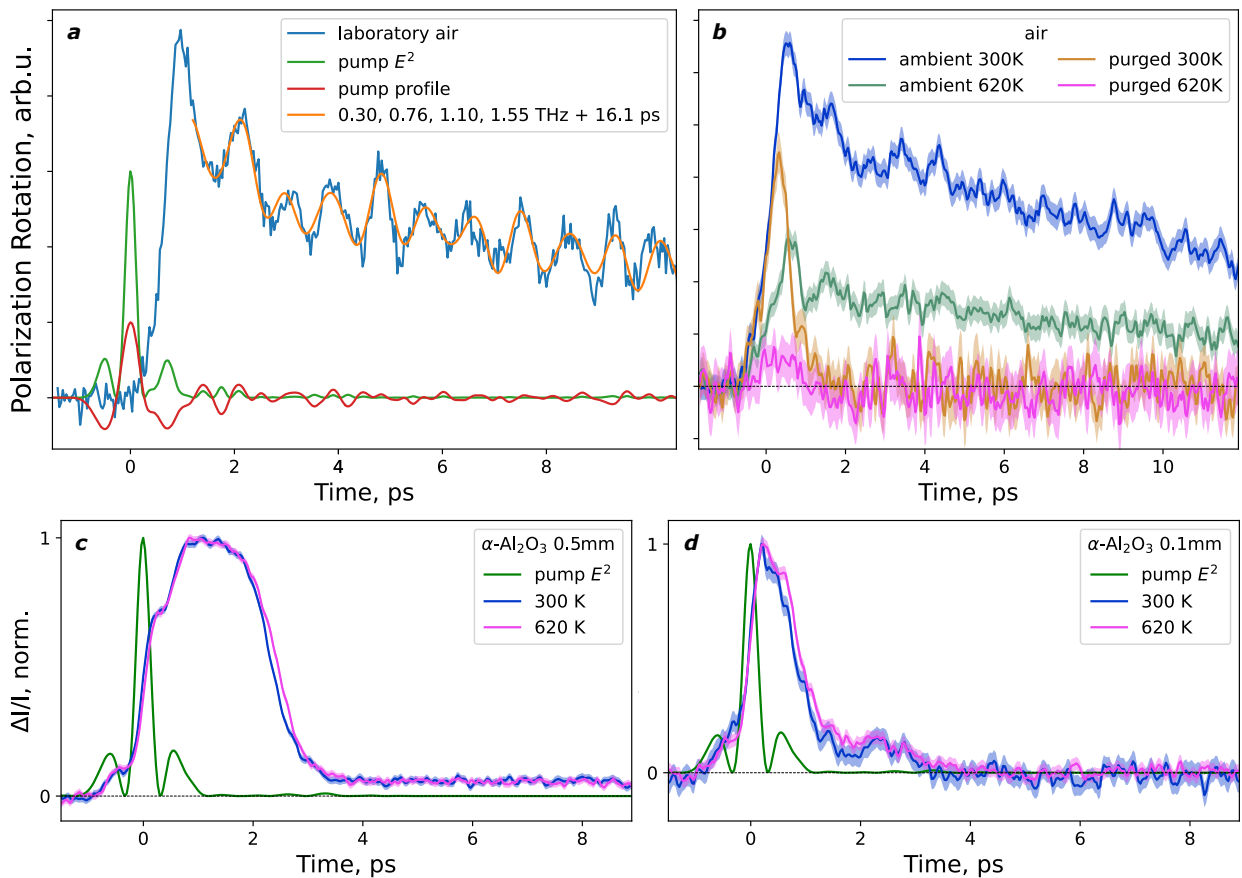

**Figure S1 | Terahertz Kerr effect (TKE) in control samples.** (**a**) TKE signal of ambient air (blue), and linear prediction fitting of oscillatory components (orange). (**b**) TKE signals of ambient air at 300 K (blue), purged air at 300 K (ochre), and ambient and purged air at 620 K (green and pink, respectively). The cited temperature is the temperature of the heating stage, with the pump and probe pulses passing through an opening in it. (**c**) TKE of 0.5 mm thick sapphire at 300 K (blue) and 620 K (pink). (**d**) TKE of 0.1 mm thick sapphire at 300 K (blue) and 620 K (pink). In (**b-d**), the shaded regions represent ±1 s.e. of the mean.

Figure S1 shows TKE measurements of ambient air and sapphire samples. The TKE of ambient air shows a signal with components matching the absorption lines of water vapor at ≈0.76 THz and ≈1.1 THz[28]. This signal decreases by ≈half when the absolute temperature is ≈doubled (Figure S1b), and disappears entirely when the volume of pump-probe spatial overlap is purged with dry nitrogen, independent of temperature (Figure S1b). We therefore assign this signal to the rotational coherences of gas-phase atmospheric water. The strength of the rotational coherence of atmospheric nitrogen at ≈8.4 ps (Figures 2a and 3a, main text) also decreases by approximately half, relative to the remainder of the signal, which remains approximately constant in magnitude, upon heating from 300 K to 620 K. This is consistent with previous measurements of rotational coherences in gas-phase molecules[29].

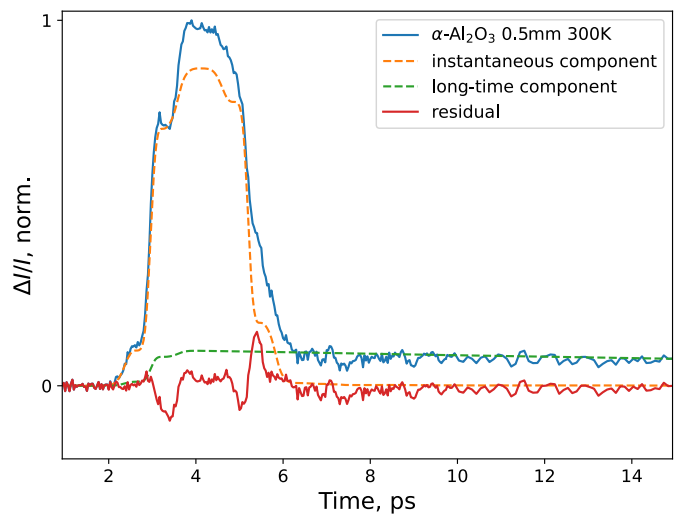

**Figure S2 | TKE signal from 0.5-mm sapphire (0001) at 300 K.** The TKE signal (blue) is fit to the sum of an instantaneous component (orange) and a long-time component that appears simultaneously. The residual is plotted in red. The rotational coherence of atmospheric nitrogen is at ≈11.4 ps with the peak pump field at ≈3 ps.

The TKE signals from sapphire samples (Figure S1cd) measured in purged atmosphere are temperature-independent. The sapphire signal is modeled (Figure S2) as a combination of an instantaneous component (Supplementary Note 1 and Ref. [24]) and a long-time component that is absent in the 0.1-mm sample (Figure S1d). Since the long-time components ("tails") in the TKE signals of $\beta$-aluminas are, unlike for sapphire, thickness-independent, but ion- and temperature-dependent, their origin must be distinct from that of the long-time sapphire signal. A more exact model of the sapphire signals remains possible, but the simpler one suffices here to distinguish it from the responses of $\beta$-aluminas.

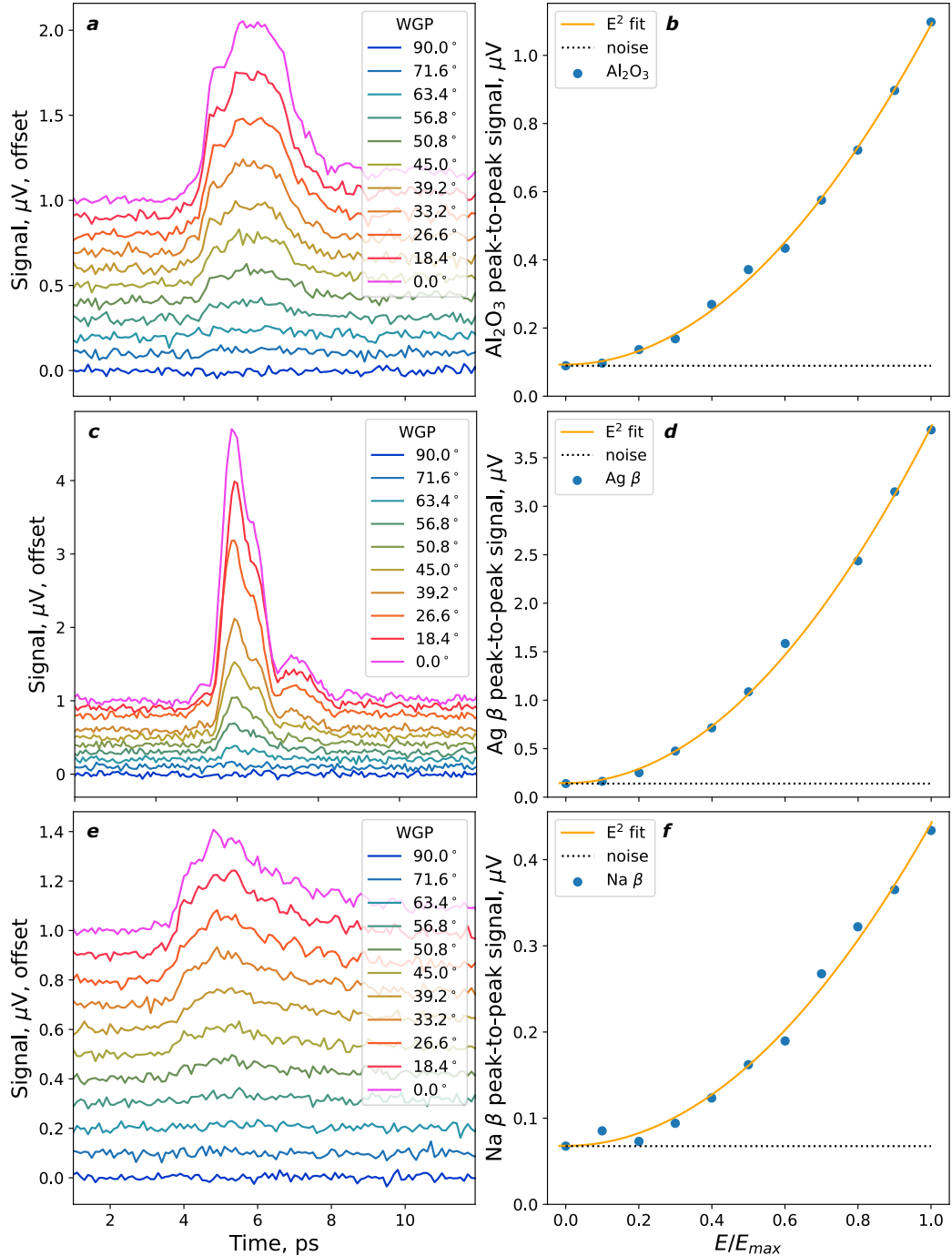

**Figure S3 | Pump field dependence of TKE signals.** Time-domain TKE signals as functions of wire-grid polarizers (WGP) angle for 0.5-mm sapphire (**a**), ≈0.4-mm Ag $\beta$-alumina (**c**), and ≈0.4-mm Na $\beta$-alumina (**e**). The pump field is maximum at the zero degrees position of the WGP, and the pump is fully blocked at 90 degrees. The peak-to-peak signals for sapphire (**b**), Ag $\beta$-alumina (**d**), and Na $\beta$-alumina (**f**) with fits (orange) to a noise level (dotted) and the intensity of the THz pump field.

All TKE signals scale with the peak intensity of the THz pump field (Figure S3). Whenever they are measurable, the kinetics of the relaxation are independent of the pump field strength.

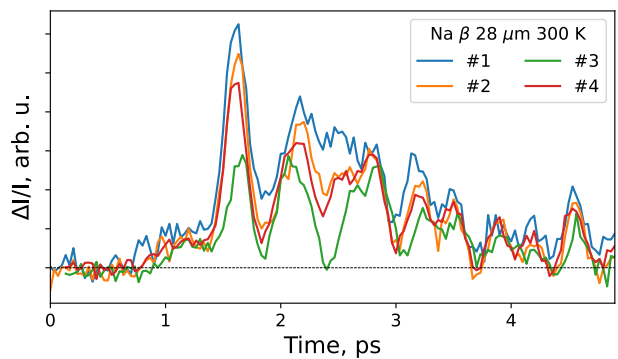

**Figure S4 | Stability of the TKE signal in Na *β*-alumina.** TKE in a thin sample of Na *β*-alumina, measured four times (#1 through #4 chronologically) over ≈4 days with other measurements in between. There was a pause between the third and fourth measurements. Here, the peak pump field is applied at 1.5 ps.

The TKE signals in *β*-aluminas show deviations at short time delays upon prolonged (several days) measurements (Figure S4), which can be reversed by stopping measurements. The measurements presented in the main text are taken when such effects are minimized, e.g., on fresh samples. The changes over the course of each measurement, up to ≈12 hours, are small, and the long-time-delay components of the signals are unaffected.

## Supplementary Note 3: Optical Conductivity from Terahertz Transmission

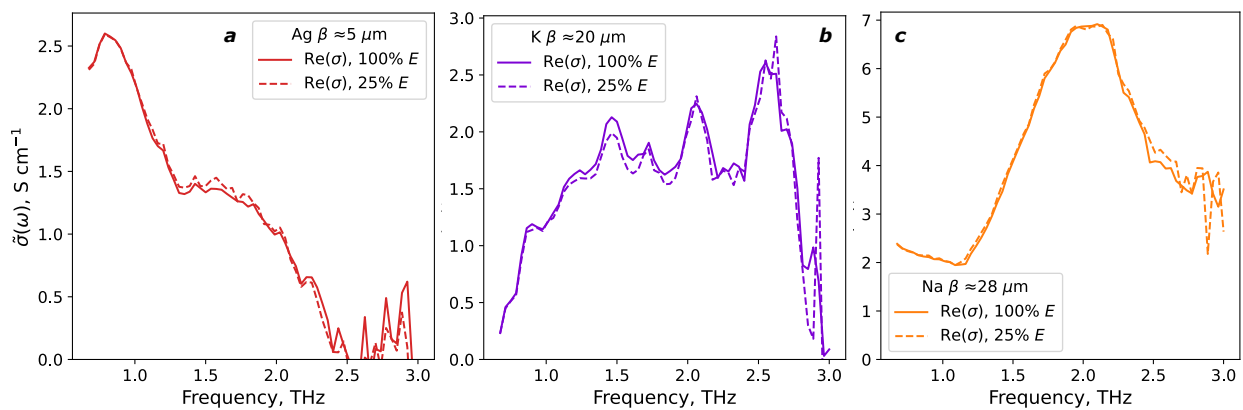

**Figure S5 | Optical conductivity of *β*-aluminas in the far infrared**. Real part of the optical conductivity σ, calculated from terahertz transmission through single-crystalline Ag (**a**, dark red), K (**b**, purple), and Na (**c**, orange) *β*-aluminas. The thicknesses of samples used are noted in the legends. The relative uncertainties in the absolute values due to the choice of starting fit guess are ≤50%. For each, the transmission measurements were performed at ≥500 kV/cm (denoted as 100% *E* field, solid lines) and a quarter of that (denoted as 25% *E* field, dashed lines). In all cases, the polarization of the pump field was parallel to the conduction planes of the crystals. Below ≈0.7 THz, and above ≈2.8 THz, the transmission was weak and conductivity calculations noisy.

The complex refractive index ñ(ω) was fit to the transmission at each frequency[6,7,30]. For a free-standing slab of thickness L, the transmission T(ω) relative to a pinhole reference is given as:

$$T(\omega) = \frac{4\tilde{n}(\omega)}{(\tilde{n}(\omega) + 1)^2} \exp(ikL(\tilde{n}(\omega) - 1))$$

Here, $k$ is the wavenumber equal to ω/c. This nonlinear equation was solved for ñ(ω) at each frequency provided an initial guess $\tilde{n}_0(\omega)$; the values shown in Figure S5 use $\tilde{n}_0$ = 2+0.5$i$. The relative uncertainty due to choices of $\tilde{n}_0$ is ≤50%, but the solver is robust to small uncertainties in other inputs, such as possible variation or e.g. 20% error in the sample thickness. This method also avoids the simplifying single-pass assumption, which may not apply at all frequencies due to strong variation in absorption across the pump spectrum.

The optical conductivities of β-aluminas show features and absolute magnitudes consistent with literature spectra of melt-grown crystals[31]: ≈0.8 THz in Ag β-alumina, 2.1 THz in Na β-alumina, and several modes between 1.5, 2.0, and 2.7 THz in K β-alumina. More importantly, for all conditions and all samples the calculated refractive index and conductivity are independent of the terahertz field for the same initial $\tilde{n}_0$ used to solve the nonlinear equation. Unlike in recent work[32], no field-strength-dependent shifting or bleaching of any features are observed over multiple measurements of each sample.

## Supplementary Note 4: Terahertz Pumps in Molecular Dynamics

The selective excitation of mobile-ion motions in β-aluminas is verified with simulations including the pump field. During the simulated terahertz pump pulse, the temperature of the mobile ions, computed by LAMMPS using the kinetic energy of the ions, rises rapidly (Figure S6ac). After the pulse is turned off, the temperature of the mobile ions begins to decrease, and the temperature of the host lattice species increases. For all materials, the terahertz pulses selectively deposit energy into the mobile ions, which then thermalize with the lattice on a picosecond timescale. The timescale of this thermalization is longer than that of the relaxation of anisotropy in hopping, highlighting the selectivity of the nonlinear optical measurement to the directional alignment of ionic hopping.

The excitation of vibrational modes by the pump pulse, in addition to hopping, is verified by tracking the displacements of the center of mass of the mobile ions, projected onto the direction of the field, and referenced to the host-lattice displacement[33]. Simplifying the ensembles of mobile ions and the host lattice to two point charges, the emitted field from coherent oscillations excited by the pump should be proportional to the velocity of the relative center-of-mass motions. This yields several material-specific vibrations(Figure S7) with frequencies typically within 10-20% of known literature infrared- and Raman-active vibrations[31,34–40]. For example, simulated vibrations are 0.6 and 1.0 THz in Ag β-alumina, and 1.7 THz in Na β-alumina. For simulated K β-alumina, the vibration at ≈1.3 THz is only excited at 300 K, while the vibration at ≈2.0 THz is excited at both 300 K and 600 K, in agreement with the TKE experiment (main text Figure 2b). For a classical simulation that does not account for partial covalency, this is excellent agreement. The simulated coherent displacements of the mobile-ion center of mass are between

5-15 picometers during the application of the pulse, in line with terahertz-frequency pump-probe studies in solid-state materials[20,21,41].

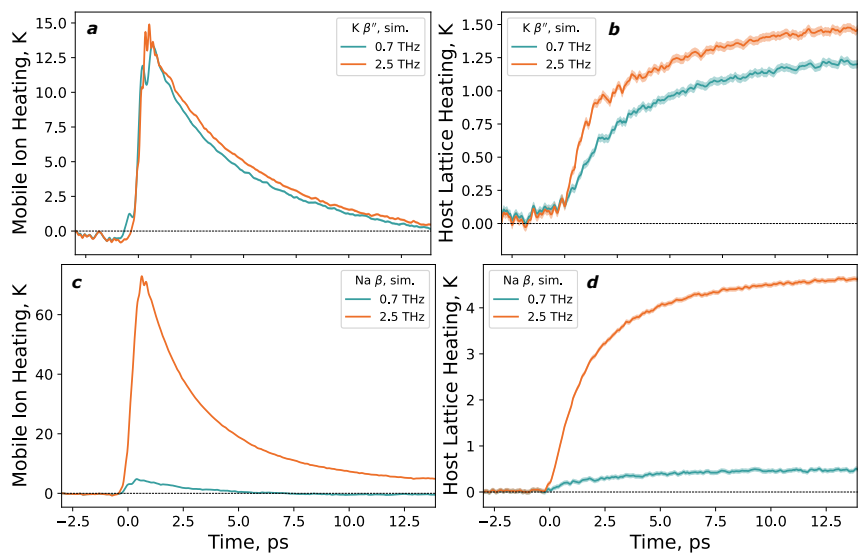

**Figure S6 | Heating of $\beta''$- and $\beta$-aluminas by simulated terahertz pulses.** (a,b) Simulated temperature rise in K $\beta''$-alumina: mobile ions (a) and the host lattice (b) with electric field pulses at 0.7 THz (teal) and 2.5 THz (orange). (c,d) Simulated temperature rise in Na $\beta$-alumina: mobile ions (c) and the host lattice (d) with electric field pulses at 0.7 THz (teal) and 2.5 THz (orange). The shaded areas are ±1 s.e. of the mean.

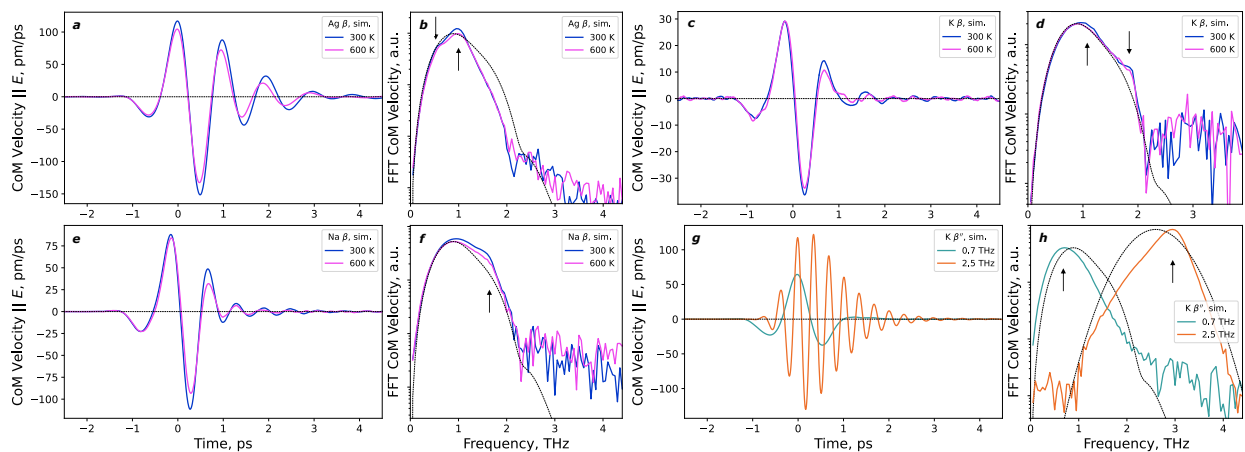

**Figure S7 | Simulated excitation of coherent motions of mobile ions by terahertz pumps**. Velocities of the center-of-mass of all mobile ions (**a,c,e,g**), and their Fourier transforms (**b,d,f,h**). Dashed black lines in the Fourier transform figures are spectra of the applied pump pulses. Highlighted vibrations: 0.6 and 1.0 THz in Ag $\beta$-alumina (**a,b**), 1.3 and 2.0 THz in K $\beta$-alumina (**c,d**), 1.7 THz in Na $\beta$-alumina (**e,f**), and 3.0 THz in K $\beta''$-alumina (**g,h**).

The *in silico* TKE of $\beta$-aluminas is dis-aggregated by the chemical environment of the mobile ions: (a) ions bound in a defect cluster, and (b) ions that are free to diffuse. The cluster consists of four mobile ions on six crystallographic lattice sites around an oxygen interstitial[10,42,43]. These ions shift between the sites within the cluster. The hopping components of *in silico* TKE of Na $\beta$-alumina (Figure S8) are similar at short time delays (Figure S8bc) – but only the un-bound ions yield anisotropic hopping past 1 ps (Figure S8b). This response is also weakly thermally activated.

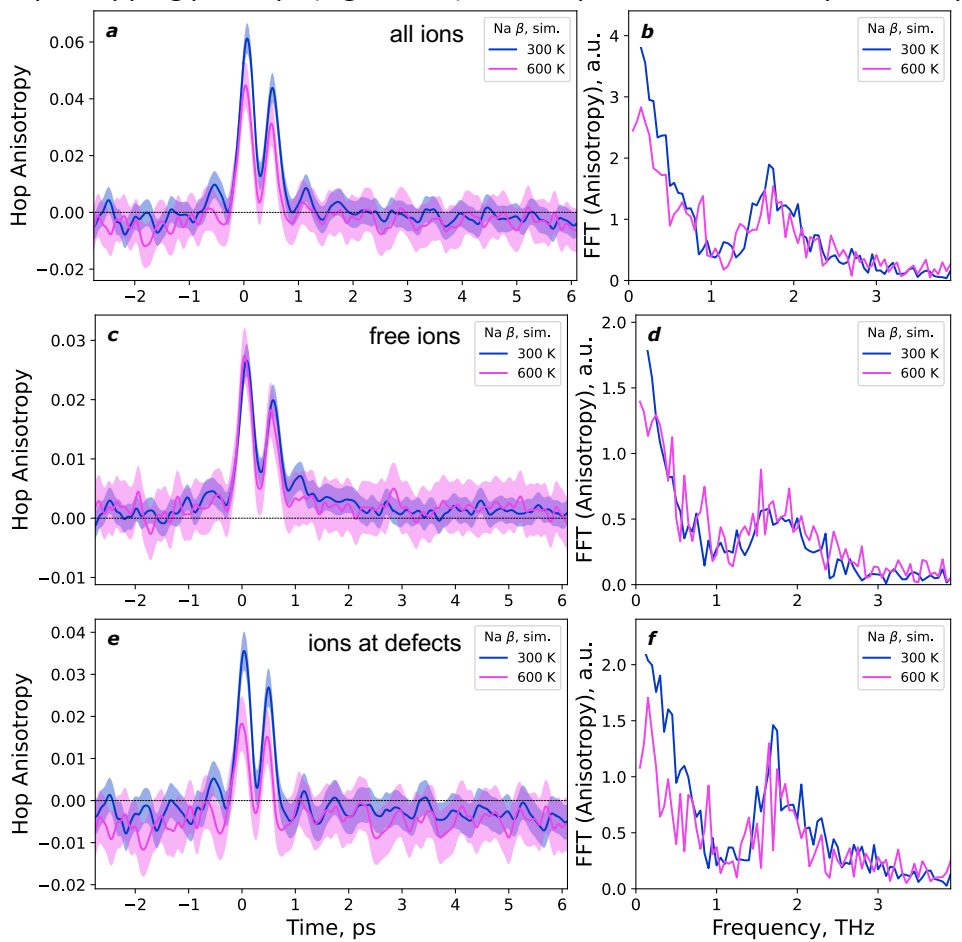

**Figure S8 | Temperature dependence of the simulated hopping anisotropy in Na $\beta$-alumina**. (**ab**) all mobile ions, (**cd**) mobile ions away from defects, i.e. "free", and (**ef**) mobile ions within defect clusters. The shaded areas are ±1 s.e. of the mean.

The temperature-independence of the hopping anisotropy response due to cluster-bound ions complicates the interpretation of experimental TKE measurements of Na and Ag $\beta$-aluminas (Figure 1cd). Additionally, the relaxation time constants are faster than in K $\beta$-alumina (Extended Data Figures 3-5). By analogy with high-frequency NMR relaxation[44,45], the temperature activation of the back-hopping in Na and Ag $\beta$-aluminas is expected to be weaker than in K $\beta$-alumina. We focus on the K $\beta$-alumina signals because they possess stronger activation, more distinguishable time constants, and simpler interpretation due to the lack of a simulated response from the defect clusters in that material. At the same time, the agreement between the classical molecular dynamics simulation of K $\beta$-alumina and the low-frequency conductivity is the worst of all $\beta$-aluminas[10]. We believe this is due to the inability of the simulation to accurately describe the two-coordinate K environment at the anti-Beevers-Ross sites, which leads to an overestimation of activation energy and overall lower conductivity. This is consistent with the pumped molecular dynamics simulations also under-estimating the lifetimes of ions in high-energy sites, and predicting a faster than experimental relaxation (Figure 3). However, the qualitative experiment-simulation agreement spans the existence, temperature-, frequency- and material-dependences. A more quantitative correspondence between simulation and experiment will require *ab initio* methods.

## Supplementary Note 5: Logical and Thermodynamic Reversibility

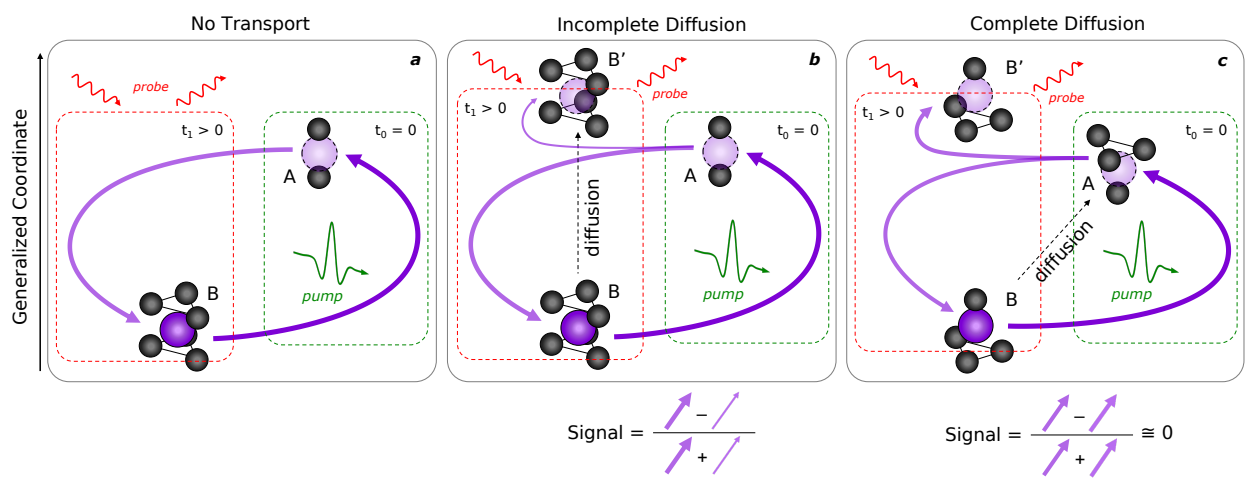

**Figure S9 | Dissipation of Pump Energy to Drive Transport**. **a**, Ions are driven from site B to a different site A, and return without coupling to transport. **b**, Ions are driven as in **a**, and some hop on to an equivalent site B', partially completing a step of transport. **c**, Ions are driven from site B to an equivalent site A, completing a step of random-walk transport; subsequent hops to B or B' are random.

Here, we draw a brief analogy to concepts and experiments in stochastic thermodynamics. The parallels between memory, correlations, and the thermodynamics of information have been formulated in multiple frameworks, and we direct the reader to several reviews[46–49]. In Figure S9, we highlight three scenarios for the dissipation of terahertz pump energy that drives ionic hopping with varying degree of coupling to transport.

In the first case, the pump drives hopping between two lattice sites (illustrated as B → A, Figure S9a). However, the subsequent hopping only returns the ions to their original sites via reverse hops. In this case, no diffusion happens. The macroscopic experiment constitutes an ensemble average over many realizations of driving a particle within a double-well potential coupled to a thermal bath. The pump performs work on the system to drive this process in finite time and surmount kinetic barriers and possibly the thermodynamic energy difference between the sites. In the second step, the system completes the reverse process: the energy imparted by the pump is thermalized to the bath. The sequence B → A → B comprises a thermodynamic cycle between two states. However, the ensemble measurement of the bulk sample is not equivalent to an average over individual ionic paths that is necessary to apply Crooks' theorem[50–52] or the Jarzynski equality[52–54]. While our experiment probes the rates of hopping following the pump, and not the work done by the system, the broad and non-resonant nature of this distribution is consistent with a distribution of work values.

In the second case (Figure S9b), the system relaxes via two competing pathways, A → B as above, and A → B' to a site B' that is crystallographically equivalent to B but in a neighboring unit cell of the crystal. This introduces coupling between sites as potential-energy wells, and allows for the pump to ultimately drive net displacement and diffusion. Since B and B' are constructed as crystallographically equivalent, the path B → A → B' constitutes diffusion (Figure S9b), analogous to the turn of a ratchet. To measure work, sites B and B' must be distinguished, such by imposing ion-permeable electrical contacts.

Assuming ideal contacts, the paths B → A → B and B → A → B' differ in the net current extractable from the pump pulse: only the latter yields a net displacement of ions and therefore current. While both pathways thermalize the pump pulse when averaged over the sample (thermodynamic irreversibility), the degree of freedom that characterizes net work (B vs B') is modified only for a fraction of the ions, with the success rate of the pathway B → A → B'. This parallels the incomplete erasure of a memory bit if the work input is too low[55–57] and suggests that higher pump fields may drive B → A → B' directly. The B → A → B' pathway mediates logical irreversibility: following the experiment, an ion that started at B' and did not move is indistinguishable from one that started at B and underwent a net displacement. By contrast, ions driven to site A by the pump are, for a time, distinguishable from the other ions at A: they are more likely to hop in a particular direction (namely, back to B) rather than randomly. In other words, for an ion at A, position alone is insufficient to determine subsequent dynamics. This insufficiency establishes a correlated hopping mechanism and distinguishes transport in this case from a random walk. The correlation between subsequent hopping and the pump-driven B → A hop is information that dissipates over time.

The TKE experiment measures velocities, not positions, and alone does not quantify the numbers of ions that follow the two paths, only their difference. In absence of ion-permeable contacts, and with the diameter of the pump pulse (≈0.5 mm) larger than that of the probe (<0.1 mm), the long-range restoring force that eventually reverses B' to B is slow and not measurable with TKE. Therefore TKE does not directly quantify the success rate for information erasure. We rely on molecular dynamics simulations and on the $\beta''$-alumina experiment (Figure 4) to distinguish the $\beta$-alumina experiments, represented by the case of Figure S9b, from the fully reversible Figure S9a. This is done by verifying that some pump-driven hops indeed contribute to the B → A → B' pathway.

The third case (Figure S9c) holds if lattice sites are equivalent, and ordering is absent. Then one hop, e.g., B → A in Figure S9c, can constitute the complete diffusion process, if velocity or mean-square-displacement correlations do not carry additional information and decay on the timescale of the pump pulse. This is the case for in K $\beta''$-alumina at 600 K, which can be represented by a random walk. In this case, the probabilities of the pathways B → A → B and B → A → B' (in Figure S9c) are the same following the pump pulse as without it, the pump-driven hops are sufficient for diffusion, and no long-time TKE signal is possible. Unlike in Figure S9b, here the ions undergoing pump-driven hops B → A are already indistinguishable from those that started at A. In this case, pump-driven hops constitute DC-limit current if perfect contacts were available to measure it.

As the $\beta$- and $\beta''$-aluminas have been the classical model systems of solid-state ionics, a variety of high-frequency measurements of electrical[58–60], optical[61–65], scattering[40,66,67], and spin[44,45,68–70] responses have been carried out. These have been partially systematized by Kamishima[63]: the activation energy increases with decreasing frequency. Our simulations[10] and experiments (this work) highlight a corresponding timescale-dependence of the correlations in ionic hopping: the complete process of conduction is measured only at timescales slower than the persistence of such correlations. Our measurements here suggest that this effect is not due to distributions of frozen potential-energy barriers in a dilute-carrier system, but to the intrinsically fluctuating stochastic dynamics of ion motions in a non-dilute, interacting one.

## Supplementary Note 6: Attempt Frequencies in Na $\beta$-alumina

Here, we seek to understand the fundamental initiation of ion transport, the attempt frequency $v_0$ for ionic hopping. In the random-walk framework (full entropy produced with every hop), $v_0$ is not an atomistic vibration, but an effective quantity with dimensions of frequency relating to the vibrational states at the saddle point of a generalized potential energy surface[71–73], which together with the difficulty of measurement has led to the use of averages over the density of states for its proxies[74–77]. However, even for a material with a simple power spectrum, the vibrational modes $Q_i$ may couple to hopping rates $H$ unequally (main text and Figure 4c). The response of $\beta$-aluminas to electrical fields driving hops is more complex than that of K $\beta''$-alumina (Figure 4c) due to the presence of defect clusters and non-equivalent crystallographic sites (Figure 1a). We first analyze steady-state simulations, followed by *in silico* TKE molecular dynamics second. We focus on the hops into the high-energy anti-Beevers-Ross (aBR) sites from the low-energy Beevers-Ross (BR) sites. As chemical doping furthermore creates distinct chemical environments for the mobile ions, here we simulate the compositional $Na_{1+2x}Al_{11}O_{17+x}$ series for $Na_2O$ doping x=0,0.01,0.1, where the practical material corresponds to x≈0.1.

For a molecular dynamics simulation, we partition the trajectories of mobile ions (typically, 100 ns for steady-state simulations) into hopping events, defined as the time point when an ion migrates between crystallographic lattice sites, and the residence times between them. A residence time is defined as the period of time that an ion spends at a lattice site following a hop into that site[10]. Here, we quantify the distributions of hopping residence times in the short-time, picosecond regime. If an ion is found to preferentially commit a hop for some (short) residence times corresponding to a multiple of some well-defined period, then such a period is empirically the inverse of the attempt frequency for hopping. Similarly, for any crystallographic site within the material, the times that elapse between an ion leaving it, and a new ion entering the site, "filling times" for short, can be analyzed with this statistical method.

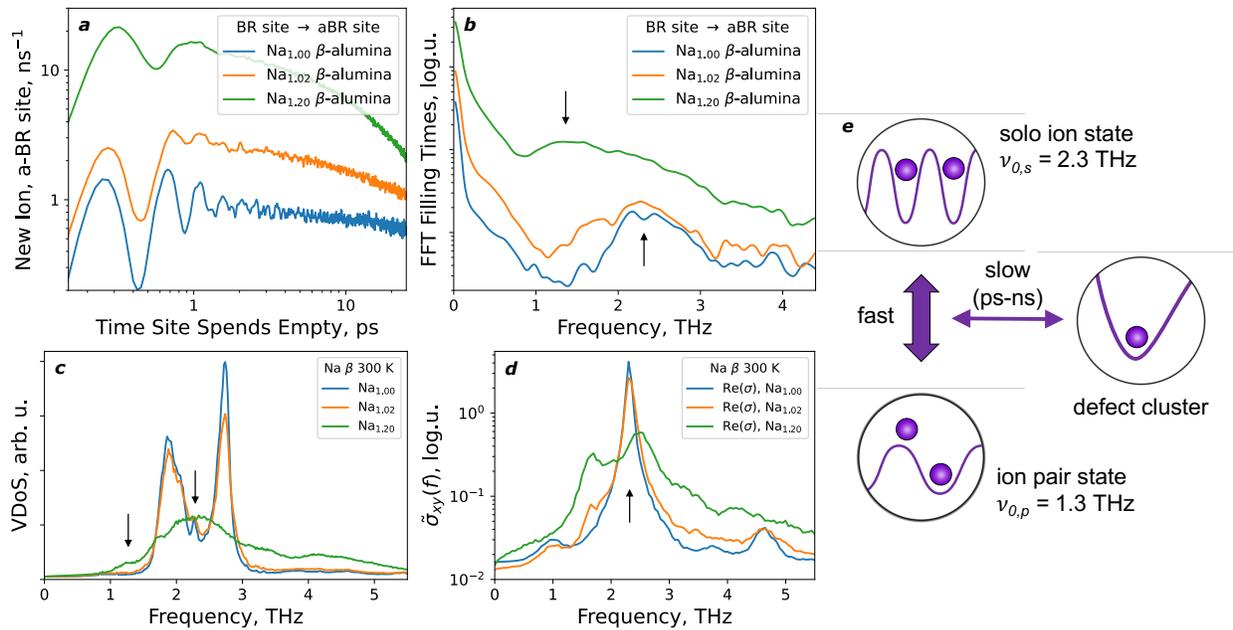

**Figure S10 | Attempt frequencies for Na $\beta$-alumina from MD simulations.** (**a**) Distributions of times that anti-Beevers-Ross sites spend empty following a Na ion hopping away, simulated at 600 K for x=0,0.01,0.1 in stoichiometries $Na_{1+2x}Al_{11}O_{17+x}$ (blue, orange, and green, respectively). (**b**) Fourier transforms of the distributions in (**a**), smoothed with a Gaussian filter of 0.1 THz st. dev. Arrows highlight peaks at 2.3 THz and 1.3 THz corresponding to the short-time structure of the distributions in (**a**). (**c**) Na vibrational density of states and (**d**) real part of the optical conductivity for the three simulated stoichiometries of Na $\beta$-alumina, with arrows highlighting the same frequencies as in (**b**). (**e**) Model of the three internal states for mobile ions in Na $\beta$-alumina: solo as in the defect-free (x=0) material, part of an mobile-ion pair, and part of a defect cluster. The first two states possess distinct attempt frequencies for hopping, 2.3 THz and 1.3 THz, respectively, whereas the defect cluster is non-diffusing.

For the defect-free material, x=0, distributions of "filling times" for high-energy aBR sites (Figure S10a, blue) show preferential hops into the sites at periods of time corresponding to 2.3 THz (Figure S10b, blue), which is a minuscule yet distinct part of the Na vibrational density of states (Figure S10c, blue) and the main feature of the infrared conductivity (Figure S10d, blue). We conclude that 2.3 THz is the attempt frequency for BR → aBR hopping in absence of neighboring Na ions already in aBR sites. This matches the experimentally measured infrared-active vibration at 2.0-2.1 THz (Supplementary Note 3). Adding even a single defect cluster per 100 formula units (x=0.01) increases the rates of hopping and already changes the frequency makeup of the hopping attempts towards lower frequencies (Figure S10b, orange). The additional Na ion locates on the high-energy sites, and necessarily perturbs the energetics of its neighbors, resulting in interstitialcy diffusion[78] that originates with perturbed, softer attempt frequencies due to the repulsion between mobile ions. For the practical simulated stoichiometry, x=0.1, the distribution of site filling times shows a longer period at short times (Figure S10a, green), which corresponds to the practical attempt frequency of 1.3 THz. This vibration is the attempt frequency for interstitialcy knock-on hopping in the practical Na $\beta$-alumina, with the simulated frequency corresponding to one measurable with inelastic neutron scattering at 1.2 THz[66]. The simulated vibrations at 1.3 and 2.3 THz (1.2 THz and 2.1 THz experimentally), are attempt frequencies for hopping from two distinct states (Figure S10e) via two mechanisms: interstitialcy ion-pair and solo, respectively. While both are present in the real material, most transport occurs via the interstitialcy pathway.

Both vibrations are distinct from the LO phonon mode at 3.0 THz detectable by TKE (Figure 2), OKE (Extended Data Figure 6), and neutron scattering[66]. The LO phonon was previously thought to correlate to hopping and activation energy[79–81]. The 1.3 THz attempt frequency corresponds to the interstitialcy nearest-neighbor mobile-ion pair, i.e., a defect created by chemical doping, whereas the 2.3 THz frequency is the attempt frequency for BR $\to$ aBR hopping in absence of nearby Na ions already in aBR sites (Figure S10e). This softening of the attempt frequency with doping demonstrates the role of repulsions between mobile ions in driving ionic conductivity. Indeed, chemical doping alters not only the concentrations of the active species, but also the very energetics of each hop. Finally, the practical attempt frequency of 1.3 THz comprises only a minor part of both the Na vibrational density of states (Figure S10c) and simulated Na optical conductivity (Figure S10d). By contrast, the optical phonon at 3.0 THz does not contribute to hopping, the same way that higher-frequency modes in K $\beta$"-alumina also do not contribute (main text, Figure 4c). Overall, our statistical analysis enables a detailed examination of the vibrational contributions to hopping beyond average descriptors such as the mobile-ion phonon band center[76,77].

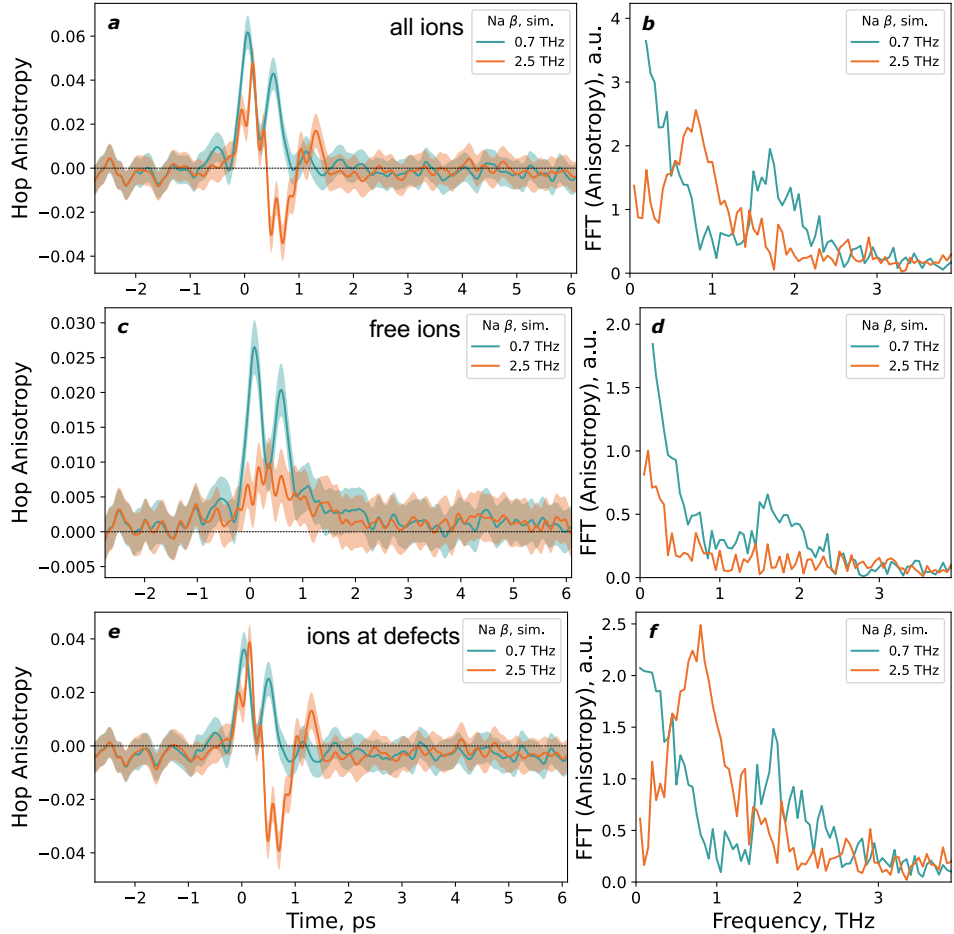

**Figure S11 | Simulated anisotropy of hopping in Na $\beta$-alumina at 300 K**. Time traces with the peak applied fields at zero time (**ace**), and Fourier transforms of the hopping anisotropy (**bdf**). Two pump pulses are compared: 0.7 THz (teal) repeated from Figure S8, and 2.5 THz (orange). The full response (**ab**) is dis-aggregated to the response of unbound ions (**cd**) and ions bound within defect clusters (**ef**). The shaded areas are ±1 s.e. of the mean.

We now simulate an *in silico* TKE experiment for Na $\beta$-alumina, at the practical stoichiometry x=0.1 and 300 K. Given two distinct mechanisms of hopping (Figure S10e) with two attempt frequencies, each could yield hopping when pumped. As for K $\beta''$-alumina (main text Figure 4c), pump field frequencies of 0.7 THz and 2.5 THz are simulated at peak fields 300 kV/cm inside the material. The simulated 2.5-THz pulse raises the temperature of the system by ≈4.5 K, 10x more than the 0.7 THz pulse at ≈0.45K (Figure S6cd). The hopping anisotropy in response to the simulated pumps (Figure S11) depends strongly both on the pumping frequency and on the location of the mobile ion (cluster-bound vs free). The response of free ions to the simulated 2.5-THz pulse is distinct from zero, but relatively weak given the strong absorption via the 2.3 THz mode. This is consistent with the solo hopping mechanism with the 2.3-THz attempt frequency providing only a minor contribution in the practical material. Finally, the overall hopping anisotropy response is strongly convoluted by defect clusters (Figure S11c). Overall, our simulation shows that TKE results should be interpreted with care for materials with multiple possible hopping mechanisms and internal states. Knowledge of the local environments of mobile ions is required to rigorously characterize the possible attempt frequencies in ionic conductors. We focus on the K $\beta$"-alumina in the main text for simplicity.

## Supplementary References

1. Hebling, J., Yeh, K.-L., Hoffmann, M. C. & Nelson, K. A. High-power THz generation, THz nonlinear optics, and THz nonlinear spectroscopy. *IEEE J. Sel. Top. Quantum Electron.* **14**, 345–353 (2008).
2. Hoffmann, M. C. Nonlinear Terahertz Spectroscopy. in *Terahertz Spectroscopy and Imaging* vol. 171 355–388 (Springer-Verlag, 2012).
3. Zheng, X., Sinyukov, A. & Hayden, L. M. Broadband and gap-free response of a terahertz system based on a poled polymer emitter-sensor pair. *Appl. Phys. Lett.* **87**, 87–89 (2005).
4. McLaughlin, C. V., Zheng, X. & Hayden, L. M. Comparison of parallel-plate and in-plane poled polymer films for terahertz sensing. *Appl. Opt.* **46**, 6283–6290 (2007).
5. Poletayev, A. D. Ion Conduction By The Picosecond: Optical Probes and Correlations. (Stanford University, 2020).
6. Neu, J. & Schmuttenmaer, C. A. Tutorial: An introduction to terahertz time domain spectroscopy (THz-TDS). *J. Appl. Phys.* **124**, (2018).
7. Morimoto, T. *et al.* Microscopic ion migration in solid electrolytes revealed by terahertz time-domain spectroscopy. *Nat. Commun.* **10**, (2019).
8. Barkhuijsen, H., de Beer, R., Bovée, W. M. M. J. & van Ormondt, D. Retrieval of frequencies, amplitudes, damping factors, and phases from time-domain signals using a linear least-squares procedure. *J. Magn. Reson.* **61**, 465–481 (1985).
9. Plimpton, S. Fast Parallel Algorithms for Short-Range Molecular Dynamics. *J. Comput. Phys.* **117**, 1–19 (1995).
10. Poletayev, A. D., Dawson, J. A., Islam, M. S. & Lindenberg, A. M. Defect-Driven Anomalous Transport in Fast-Ion Conducting Solid Electrolytes. *https://arxiv.org/abs/2105.08761* (2021).
11. Edvardsson, S., Ojamae, L. & Thomas, J. O. A study of vibrational modes in Na+ beta - alumina by molecular dynamics simulation. *J. Phys. Condens. Matter* **6**, 1319–1332 (1994).
12. Mishra, P. K., Vendrell, O. & Santra, R. Ultrafast Energy Transfer from Solvent to Solute Induced by Subpicosecond Highly Intense THz Pulses. *J. Phys. Chem. B* **119**, 8080–8086 (2015).
13. Mishra, P. K., Bettaque, V., Vendrell, O., Santra, R. & Welsch, R. Prospects of Using High-Intensity THz Pulses to Induce Ultrafast Temperature-Jumps in Liquid Water. *J. Phys. Chem. A* **122**, 5211–5222 (2018).
14. Zalden, P. *et al.* Molecular polarizability anisotropy of liquid water revealed by terahertz-induced transient orientation. *Nat. Commun.* **9**, 1–7 (2018).
15. Elgabarty, H. *et al.* Energy transfer within the hydrogen bonding network of water following resonant terahertz excitation. *Sci. Adv.* **6**, 1–15 (2020).
16. Finneran, I. A. *et al.* 2D THz-THz-Raman Photon-Echo Spectroscopy of Molecular Vibrations in Liquid Bromoform. *J. Phys. Chem. Lett.* **8**, 4640–4644 (2017).
17. Finneran, I. A. *et al.* Coherent two-dimensional terahertz-terahertz-Raman spectroscopy.

*Proc. Natl. Acad. Sci. U. S. A.* **113**, 6857–6861 (2016).
18. Savolainen, J., Ahmed, S. & Hamm, P. Two-dimensional Raman-terahertz spectroscopy of water. *Proc. Natl. Acad. Sci. U. S. A.* **110**, 20402–20407 (2013).
19. Ciardi, G., Berger, A., Hamm, P. & Shalit, A. Signatures of Intra- And Intermolecular Vibrational Coupling in Halogenated Liquids Revealed by Two-Dimensional Raman-Terahertz Spectroscopy. *J. Phys. Chem. Lett.* **10**, 4463–4468 (2019).
20. Von Hoegen, A., Mankowsky, R., Fechner, M., Först, M. & Cavalleri, A. Probing the interatomic potential of solids with strong-field nonlinear phononics. *Nature* **555**, 79–82 (2018).
21. Kozina, M. *et al.* Terahertz-driven phonon upconversion in SrTiO3. *Nat. Phys.* **15**, 387–392 (2019).
22. Kryvohuz, M. & Mukamel, S. Multidimensional measures of response and fluctuations in stochastic dynamical systems. *Phys. Rev. A - At. Mol. Opt. Phys.* **86**, 1–12 (2012).
23. Ikeda, T., Ito, H. & Tanimura, Y. Analysis of 2D THz-Raman spectroscopy using a non-Markovian Brownian oscillator model with nonlinear system-bath interactions. *J. Chem. Phys.* **142**, (2015).
24. Sajadi, M., Wolf, M. & Kampfrath, T. Terahertz-field-induced optical birefringence in common window and substrate materials. *Opt. Express* **23**, 28985 (2015).
25. Maehrlein, S. F. *et al.* Decoding ultrafast polarization responses in lead halide perovskites by the two-dimensional optical Kerr effect. *Proc. Natl. Acad. Sci.* **118**, e2022268118 (2021).
26. Huber, L., Maehrlein, S. F., Wang, F., Liu, Y. & Zhu, X. Y. The ultrafast Kerr effect in anisotropic and dispersive media. *J. Chem. Phys.* **154**, (2021).
27. Grischkowsky, D., Keiding, S. S., van Exter, M. & Fattinger, C. Far-infrared time-domain spectroscopy with terahertz beams of dielectrics and semiconductors. *J. Opt. Soc. Am. B* **7**, 2006 (1990).
28. van Exter, M., Fattinger, C. & Grischkowsky, D. Terahertz time-domain spectroscopy of water vapor. *Opt. Lett.* **14**, 1128 (1989).
29. Fleischer, S., Zhou, Y., Field, R. W. & Nelson, K. A. Molecular orientation and alignment by intense single-cycle THz pulses. *Phys. Rev. Lett.* **107**, 1–5 (2011).
30. Spies, J. A. *et al.* Terahertz Spectroscopy of Emerging Materials. *J. Phys. Chem. C* **124**, 22335–22346 (2020).
31. Allen, S. J., Cooper, A. S., Derosa, F., Remeika, J. P. & Ulasi, S. K. Far-infrared absorption and ionic conductivity of Na, Ag, Rb, and K β-alumina. *Phys. Rev. B* **17**, 4031–4042 (1978).
32. Minami, Y. *et al.* Macroscopic Ionic Flow in a Superionic Conductor Na+ β-Alumina Driven by Single-Cycle Terahertz Pulses. *Phys. Rev. Lett.* **124**, 147401 (2020).
33. Marcolongo, A. & Marzari, N. Ionic correlations and failure of Nernst-Einstein relation in solid-state electrolytes. *Phys. Rev. Mater.* **1**, 1–4 (2017).
34. Colomban, P., Mercier, R. & Lucazeau, G. Vibrational study of and conduction mechanism in β alumina. II. Nonstoichiometric β alumina. *J. Chem. Phys.* **75**, 1388–1399 (1981).
35. Colomban, P. & Lucazeau, G. β″-and ion-rich β-alumina: Comparison of vibrational spectra and conductivity parameters. *Solid State Ionics* **2**, 277–288 (1981).
36. Hao, C. H., Chase, L. L. & Mahan, G. D. Raman scattering in beta-alumina. *Phys. Rev. B* **13**, 4306–4313 (1976).

37. Chase, L. L., Hao, C. H. & Mahan, G. D. Raman scattering from sodium and silver in β-alumina. *Solid State Commun.* **18**, 401–403 (1976).
38. Klein, P. B., Schafer, D. E. & Strom, U. Cation interstitial pair modes in the vibrational spectra of mixed β-aluminas. *Phys. Rev. B* **18**, 4411–4421 (1978).
39. Barker, A. S., Ditzenberger, J. A. & Remeika, J. P. Lattice vibrations and ion transport spectra in β-alumina. I. Infrared spectra. *Phys. Rev. B* **14**, 386–394 (1976).
40. Lucazeau, G. Infrared, Raman and neutron scattering studies of β- and β″-alumina: a static and dynamical structure analysis. *Solid State Ionics* **8**, 1–25 (1983).
41. Neugebauer, M. J. et al. Comparison of coherent phonon generation by electronic and ionic Raman scattering in $LaAlO_3$. *Phys. Rev. Res.* **3**, 13126 (2021).
42. Wolf, D. On the mechanism of diffusion in sodium beta alumina. *J. Phys. Chem. Solids* **40**, 757–773 (1979).
43. Zendejas, M. A. & Thomas, J. O. Conduction mechanisms in solid electrolytes: Na+ beta-alumina. *Phys. Scr.* **1990**, 235–244 (1990).
44. Walstedt, R. E., Dupree, R., Remeika, J. P. & Rodriguez, A. Na23 nuclear relaxation in Na β-alumina: Barrier-height distributions and the diffusion process. *Phys. Rev. B* **15**, 3442–3454 (1977).
45. Iwai, Y., Kamishima, O., Kuwata, N., Kawamura, J. & Hattori, T. 109Ag NMR and relaxation mechanism in single crystal Ag β-alumina. *Solid State Ionics* **179**, 862–866 (2008).
46. Parrondo, J. M. R., Horowitz, J. M. & Sagawa, T. Thermodynamics of information. *Nat. Phys.* **11**, 131–139 (2015).
47. Jarzynski, C. Equalities and Inequalities: Irreversibility and the Second Law of Thermodynamics at the Nanoscale. *Annu. Rev. Condens. Matter Phys.* **2**, 329–351 (2011).
48. Ciliberto, S. Experiments in Stochastic Thermodynamics: Short History and Perspectives. *Phys. Rev. X* **7**, 16–21 (2017).
49. Ciliberto, S. & Lutz, E. The Physics of Information: From Maxwell to Landauer. in *Energy Limits in Computation* 155–175 (Springer International Publishing, 2019). doi:10.1007/978-3-319-93458-7_5.
50. Crooks, G. E. Nonequilibrium measurements of free energy differences for microscopically reversible Markovian systems. *J. Stat. Phys.* **90**, 1481–1487 (1998).
51. Collin, D. et al. Verification of the Crooks fluctuation theorem and recovery of RNA folding free energies. *Nature* **437**, 231–234 (2005).
52. Liphardt, J., Dumont, S., Smith, S. B., Tinoco, I. & Bustamante, C. Equilibrium information from nonequilibrium measurements in an experimental test of Jarzynski's equality. *Science (80-. ).* **296**, 1832–1835 (2002).
53. Jarzynski, C. Nonequilibrium equality for free energy differences. *Phys. Rev. Lett.* **78**, 2690–2693 (1997).
54. Jarzynski, C. Equilibrium free-energy differences from nonequilibrium measurements: A master-equation approach. *Phys. Rev. E* **56**, 5018–5035 (1997).
55. Bérut, A. et al. Experimental verification of Landauer's principle linking information and thermodynamics. *Nature* **483**, 187–189 (2012).
56. Gavrilov, M., Chétrite, R. & Bechhoefer, J. Direct measurement of weakly nonequilibrium system entropy is consistent with Gibbs–Shannon form. *Proc. Natl. Acad. Sci. U. S. A.* **114**,

11097–11102 (2017).
57. Gavrilov, M. & Bechhoefer, J. Erasure without work in an asymmetric double-well potential. *Phys. Rev. Lett.* **117**, 29–32 (2016).
58. Barker, A. S., Ditzenberger, J. A. & Remeika, J. P. Lattice vibrations and ion transport spectra in β-alumina. II. Microwave spectra. *Phys. Rev. B* **14**, 4254–4265 (1976).
59. Hoppe, R., Kloidt, T. & Funke, K. Frequency-Dependent Conductivities of RbAg4I5 and Na-β"-Alumina from Radio to FIR Frequencies. *Berichte der Bunsengesellschaft für Phys. Chemie* **95**, 1025–1028 (1991).
60. Kamishima, O., Iwai, Y. & Kawamura, J. Small power-law dependence of ionic conductivity and diffusional dimensionality in β-alumina. *Solid State Ionics* **281**, 89–95 (2015).
61. Suemoto, T. & Ishigame, M. Quasielastic Light Scattering in superionic β-alumina. *Phys. Rev. B* **32**, 4126 (1985).
62. Kawaharada, I., Hattori, T., Ishigame, M. & Shin, S. Quasielastic light scattering in Na1-xAgxβ-aluminas. *Solid State Ionics* **69**, 79–84 (1994).
63. Kamishima, O. *et al.* Temperature dependence of low-lying phonon dephasing by ultrafast spectroscopy (optical Kerr effect) in Ag β-alumina and Tl β-alumina. *J. Phys. Condens. Matter* **19**, 456215 (2007).
64. Allen, S. J. & Remeika, J. P. Direct measurement of the attempt frequency for ion diffusion in Ag and Na β-alumina. *Phys. Rev. Lett.* **33**, 1478–1481 (1974).
65. Hayes, W. Light scattering by superionic conductors. in *Light Scattering in Solids III. Topics in Applied Physics.* vol. 51 93–120 (Springer Berlin Heidelberg, 1982).
66. McWhan, D. B., Shapiro, S. M., Remeika, J. P. & Shirane, G. Neutron-scattering studies on beta-alumina. *J. Phys. C Solid State Phys.* **8**, L487 (1975).
67. Lucazeau, G., Dohy, D., Fanjat, N. & Dianoux, A. J. Study of the dynamics of a single crystal of Na+β-Al2O3 by neutron scattering. *Solid State Ionics* **28–30**, 1611–1616 (1988).
68. Bjorkstam, J. L. & Villa, M. NMR studies of superionic β-aluminas. *J. Phys.* **42**, 345–351 (1981).
69. Bjorkstam, J. L., Villa, M. & Farrington, G. C. Temperature dependence of the Na+ distribution in β-aluminas. *Solid State Ionics* **5**, 153–156 (1981).
70. Greenbaum, S. G. & Strom, U. Low-temperature nuclear spin relaxation in β-aluminas. *Solid State Commun.* **46**, 437–440 (1983).
71. Vineyard, G. H. Frequency factors and isotope effects in solid state rate processes. *J. Phys. Chem. Solids* **3**, 121–127 (1957).
72. Kadkhodaei, S. & Davariashtiyani, A. Phonon-assisted diffusion in bcc phase of titanium and zirconium from first principles. *Phys. Rev. Mater.* **4**, 043802 (2020).
73. de Klerk, N. J. J., van der Maas, E. & Wagemaker, M. Analysis of Diffusion in Solid-State Electrolytes through MD Simulations, Improvement of the Li-Ion Conductivity in β-Li 3 PS 4 as an Example. *ACS Appl. Energy Mater.* **1**, 3230–3242 (2018).
74. Ohno, S. *et al.* Materials design of ionic conductors for solid state batteries. *Prog. Energy* **2**, 022001 (2020).
75. Muy, S., Schlem, R., Shao-Horn, Y. & Zeier, W. G. Phonon–Ion Interactions: Designing Ion Mobility Based on Lattice Dynamics. *Adv. Energy Mater.* **11**, 2002787 (2021).
76. Krauskopf, T. *et al.* Comparing the Descriptors for Investigating the Influence of Lattice

Dynamics on Ionic Transport Using the Superionic Conductor $Na_3PS_{4-x}Se_x$. *J. Am. Chem. Soc.* **140**, 14464–14473 (2018).
77. Muy, S. *et al.* Tuning mobility and stability of lithium ion conductors based on lattice dynamics. *Energy Environ. Sci.* **11**, 850–859 (2018).
78. Morgan, B. J. & Madden, P. A. Relationships between atomic diffusion mechanisms and ensemble transport coefficients in crystalline polymorphs. *Phys. Rev. Lett.* **112**, 4–6 (2014).
79. Wakamura, K. Roles of phonon amplitude and low-energy optical phonons on superionic conduction. *Phys. Rev. B - Condens. Matter Mater. Phys.* **56**, 11593–11599 (1997).
80. Wakamura, K. Origin of the low-energy mode in superionic conductors. *Phys. Rev. B - Condens. Matter Mater. Phys.* **59**, 3560–3568 (1999).
81. Bachman, J. C. *et al.* Inorganic Solid-State Electrolytes for Lithium Batteries: Mechanisms and Properties Governing Ion Conduction. *Chem. Rev.* **116**, 140–162 (2016).